\documentclass[12pt,a4paper]{article}
\title{qKZ equation and ground state of the $O(1)$ loop model with
  open boundary conditions}
\date{\today}
\usepackage[english]{babel}
\usepackage[dvips]{graphicx}
\usepackage{latexsym,amsfonts,amssymb}
\usepackage{amsmath}
\usepackage{amsthm}
\usepackage{eufrak}
\usepackage{epsf}

\usepackage{xypic}
\usepackage{amscd}
\usepackage[all]{xy}
\usepackage{textcomp} 
\usepackage{makeidx} 
\usepackage{mathrsfs} 

\def\eq{\begin{equation}}
\def\en{\end{equation}}

\def\>>{\rangle}

\def\sH{\mathscr H}

\def\bC{\mathbb C}

\def\cH{\mathcal H}

\def\Rc{\check{R}}

\newtheorem{lemma}{\emph {Lemma}}

\def\sk{\vskip .4cm}

\begin{document}	


\thispagestyle{empty}

\begin{titlepage}

\maketitle

\vspace{-0.5 truecm}

\centerline{\large Luigi
  Cantini\footnote[1]{Laboratoire de Physique Th\'eorique, Ecole
    Normale 
    Sup\'erieure, 24 rue Lhomond, 75005 Paris, France.  {\small \tt
      <luigi.cantini@ens.fr>}}
    }

\vspace{.3cm}

\vspace{1.0 cm}

\begin{abstract}
\noindent
We consider the qKZ equations based on the two boundaries Temperley
Lieb algebra. We construct their solution in the case $s=q^{-3/2}$
using a recursion relation. At the combinatorial point $q^{1/2}=
e^{-2\pi i/3}$ the solution reduces to the ground state of the dense
$O(1)$ loop model on a strip with open boundary conditions. We present
an alternative construction of such ground state based on the knowledge
of the ground state of the same model with mixed boundary conditions
and prove that the sum rule as of its components is given by the
product of four symplectic characters.

\end{abstract}

\end{titlepage}




\section{Introduction}\label{introduction}

The interplay between statistical mechanics and combinatorics has
always been of great interest both for physicists and for
mathematician. The observations of Razumov and Stroganov in their
seminal papers \cite{raz-strog1,raz-strogO(1)_1} (see also \cite{BdGN}),
and the body of work that followed, 
exemplifies an instance of such an interplay. What Razumov and
Stroganov found was that the properly normalized components of the
ground state of the 
dense $O(1)$ model enumerate classes of so called Fully Packed Loop (FPL)
configurations of given topology. The boundary conditions on the
$O(1)$ model are reflected on the symmetries of the FPL (see \cite{degier}
for further explications). An approach to these problems, that has
revealed to be particularly powerful, was initiated by Di Francesco
and Zinn-Justin \cite{pdf-pzj-1}. They used the integrable structure of the
dense $O(1)$ model and generalized it by introducing spectral
parameters, still preserving integrability. The advantage in dealing
with a more complicated problem was at the beginning a technical one:
one could exploit the richer structure coming from the polynomial
nature of the ground state of the generalized model to better
handle it. Later it was realized that the idea of Di Francesco and
Zinn-Justin opened new perspective on the problem by relating it to the
study of so called qKZ equations \cite{pdf-pzj-qKZ-1} and
affine Hecke algebras \cite{pasquier}, to 
algebraic geometry of certain affine varieties \cite{pdf-pzj-qKZ-1} and allowing to
push forward some further intriguing conjectures (in part already
proved \cite{conjecture}) relating the homogeneous specialization of the solution
of certain qKZ equation to refined enumeration of Plane Partitions. 

In the present work, we will study the qKZ equations related to the so
called two boundaries Temperley Lieb algebra \cite{mitra,
  degier_nichols, JS2}. When specialized
at $q^{1/2}= e^{-2\pi i/3}$ the solution of such a system of
equations is the ground state of the inhomogeneous $O(1)$ loop model
with so called open boundary conditions. For generic $q$ instead, the
problem is related to the study of certain Laurent polynomial
representations of the affine Hecke algebras $\cH(C_N)$ of type
$C_N$ and of the doubly affine Hecke algebras of type $C^\vee C_N$.  
The main difficulty we encounter in dealing with two
open boundaries is the absence of a completely factorized component.
The presence of such a component in the cases with other
boundary conditions, previously  studied, allowed to 
fix the unknown factor in the recursion
relations which in turn allowed to derive the sum rule at the
combinatorial point $q^=1$. Here we adopt the opposite strategy, we derive
first the 
recursion relations by a method that circumvent the full knowledge of
a component and then we use such recursions to fix the simplest
component from which all the other can be derived. 

Still we will show that at $q^{1/2}= e^{-2\pi i/3}$ the recursion
relations are not sufficient to derive the sum of the components. 
%
%
Therefore we must resort to a different derivation of the whole eigenstate, based
on mappings to systems with a single open boundary. This way we
show that the degree is preserved and we compute the sum rule, which
is given by the product of four symplectic characters,
correcting some recent claims \cite{dGPS}. 

The paper is organized as follows. In Section \ref{algebra} we
introduce the two boundaries Temperley 
Lieb algebra and the representation of this algebra on extended link
patterns. The $\Rc$ and $K$ matrices based on baxterization of the two
boundaries Temperley Lieb algebra, solution of the Yang-Baxter
equation and of the boundary Yang-Baxter equation, is presented in
Section \ref{scatt_matr}. Then, in Section \ref{qKZ-section} we
introduce the qKZ equations and explain some of the properties of their
solution. In particular in Section \ref{affine_sectio} we explain the
relation between our qKZ equations and the representation theory of
the (doubly) affine Hecke algebras of type $C$ and in Section \ref{app_rec_bulk}
we derive the recursion relations. In Section \ref{sol_section} we
concentrate on the case where the parameter $s$ of qKZ assumes the
value $s=q^{-3/2}$, we use in such a case the recursion relations to
construct the solution of qKZ. The specialization
$q^{1/2}= e^{-2\pi i/3}$ is studied in Section
\ref{specialization}, where we explain how to derive the full solution
of the qKZ equations with open boundaries from 
the solution with mixed boundary conditions, by defining certain
mappings of 
representations.

\section{The two boundaries Temperley Lieb algebra}\label{algebra}

The problem we are going to consider is based on a boundary extension of the
well known Temperley-Lieb algebra, called $2$ boundaries Temperley
Lieb algebra  \cite{mitra, degier, degier_nichols}.  
The Temperley Lieb algebra $TL_N$ can be defined as the free algebra
with generators $e_i$, for $i=1,\dots,N-1$, and relations 
$$
e_i^2 = \tau e_i ; ~~~~~~ e_i e_j = e_j e_i~~~~ {\rm for}~~~~ |i-j| > 1;
$$
$$
e_i e_{i \pm 1} e_i = e_i ~~~~~~ {\rm for }~~~~ 2 \leq i \leq N-2 ; 
$$ 
This algebra has an appealing graphical representation in terms of non crossing
link patterns connecting $N$ points on the the top and $N$ points on
the bottom of a finite strip, with the graphical rules 
\sk
\begin{center}
\includegraphics[scale=.6]{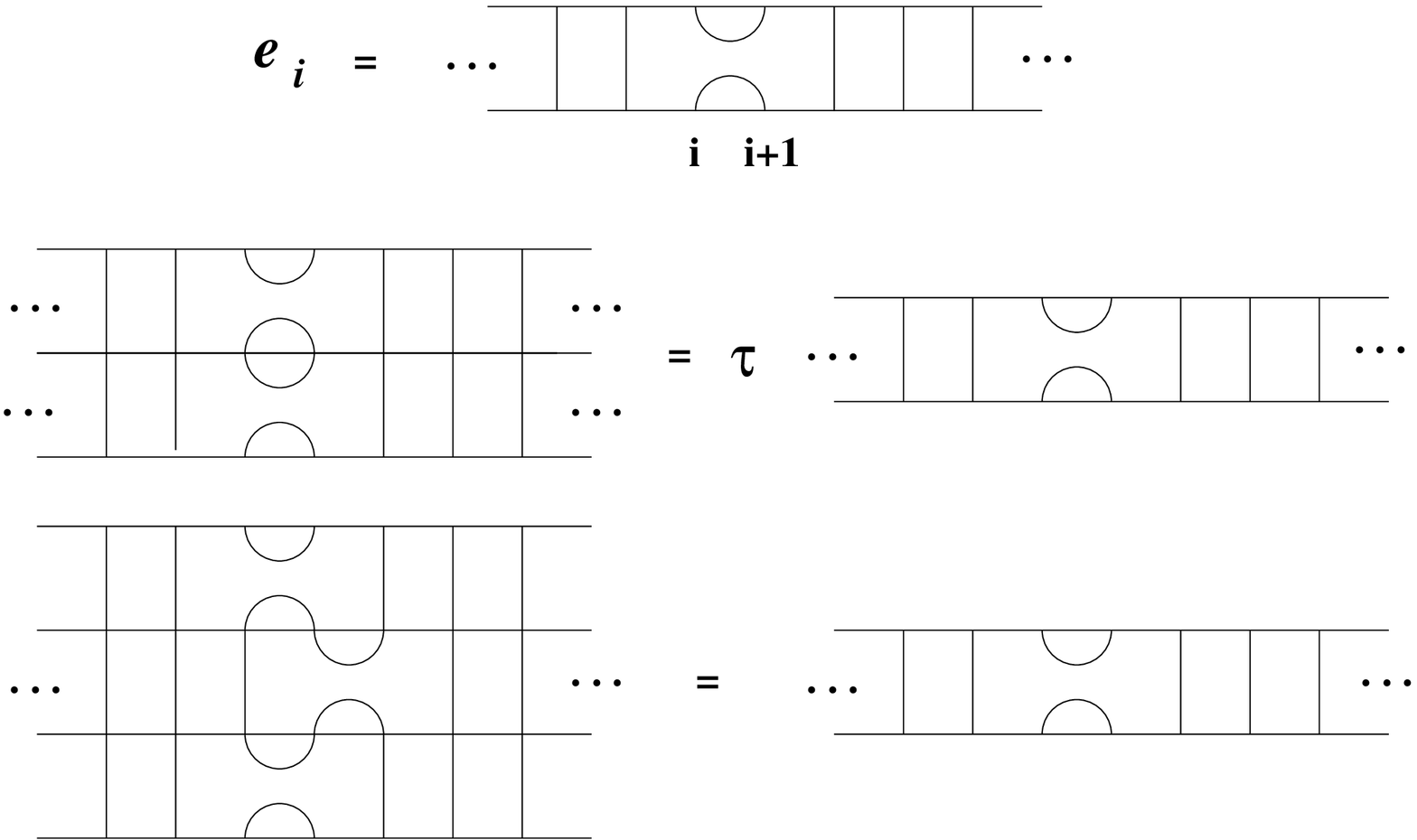}
\end{center}
\sk
A first extension of the TL algebra is obtained by adding to the
generators of $TL_N$  a so called boundary operator $f_R$ 
with the following commutation rules
$$
f_R^2 = \tau_R f_R ~~~~ e_{N-1} f_R e_{N-1} = e_{N-1} ~~~~
e_i f_R = f_R e_i ~~~~{\rm for}~~~~ i < N-1.
$$
This algebra, which is sometimes called 1BTL (one boundary
Temperley-Lieb algebra), appeared for the first time in the paper 
\cite{martin_saleur} where it 
was called blob algebra, then it was studied in different contexts
(for example \cite{JS1}). Here we
call it $TL^{(c,o)}_N$ since it is naturally related to certain
  statistical mechanics loop models having closed boundary condition
  on one side and open boundary conditions on the other side. 
Graphically the generator $f_R$ can be represented as 
\sk
\begin{center}
\includegraphics[scale=.6]{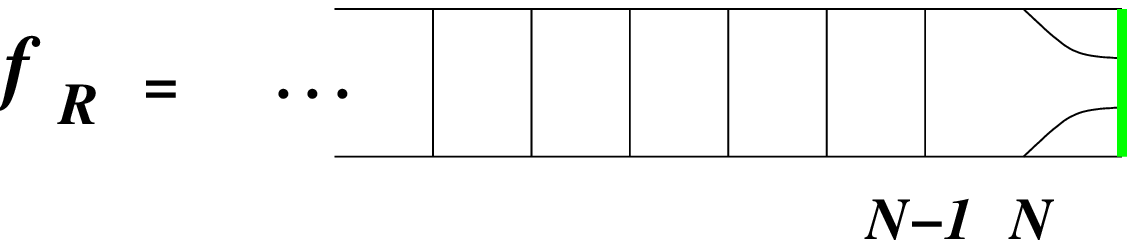}
\end{center}
and the commutation rules correspond to the following graphical
relations \sk
\begin{center}
\includegraphics[scale=.6]{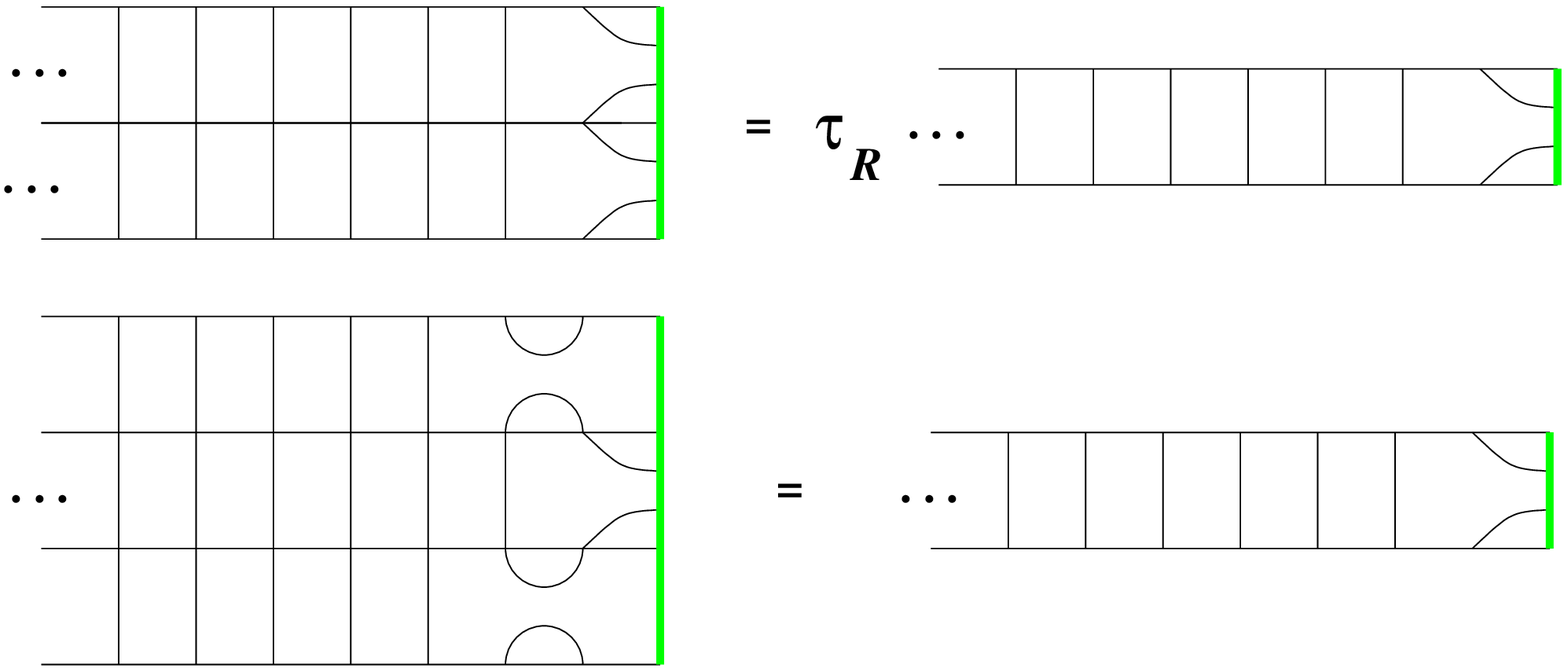}
\end{center}
The algebra obtained by adding to the generators of the Temperley-Lieb
algebra a generator $f_L$ ``to 
the left'', with commutation relation 
$$
f_L^2 = \tau_L f_L ~~~~ e_{1} f_L e_{1} = e_{1}
~~~~ e_i f_L = f_L e_i
~~~~{\rm for}~~~~ i > 1;  
$$
will be called $TL^{(o,c)}_N$. The graphical representation of the
generator $f_L$ is similar to the one of $f_R$ and the commutation
rules look as follows
\begin{center}
\includegraphics[scale=.6]{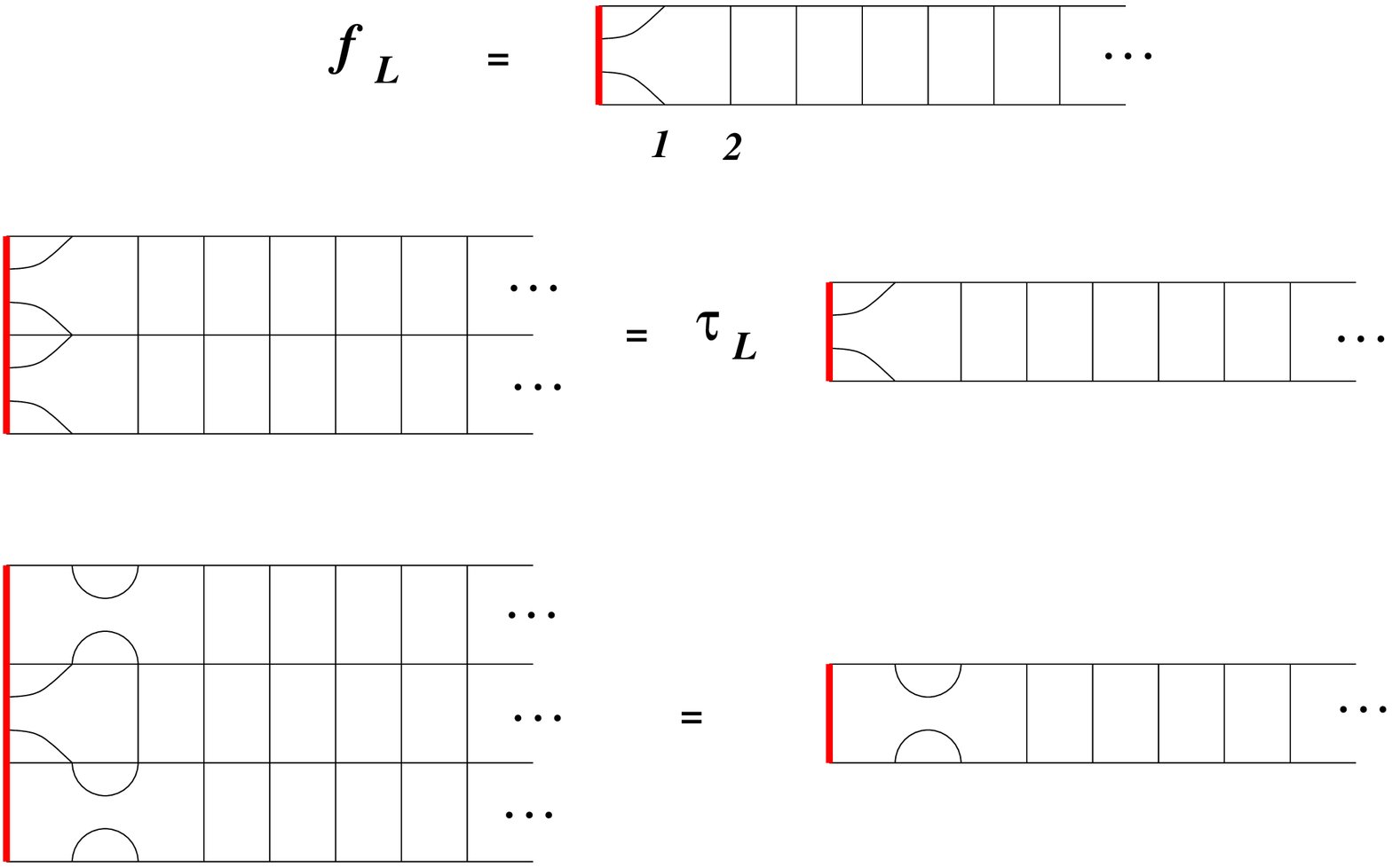}
\end{center}
In the statistical mechanics model the boundary conditions are now exchanged
  with respect to the case with $TL^{(o,c)}_N$. 

Finally we define the algebra corresponding to open boundary
conditions on both sides $TL^{(o,o)}_N$ if we add to $TL_N$ both $f_R$ and $f_L$ and require
them to commute among themselves. While the algebras $TL_N$,
$TL^{(o,c)}_N$ and $TL^{(c,o)}_N$ are finite dimensional, this is not
the case for $TL^{(o,o)}_N$. It is quite easy to understand why by
looking at the graphical representation: the commutation rules given
above do not allow to erase the lines connecting the two
boundaries. This means that for example all the element in the
following picture have to be considered as distinct. 
\begin{center}
\includegraphics[scale=.4]{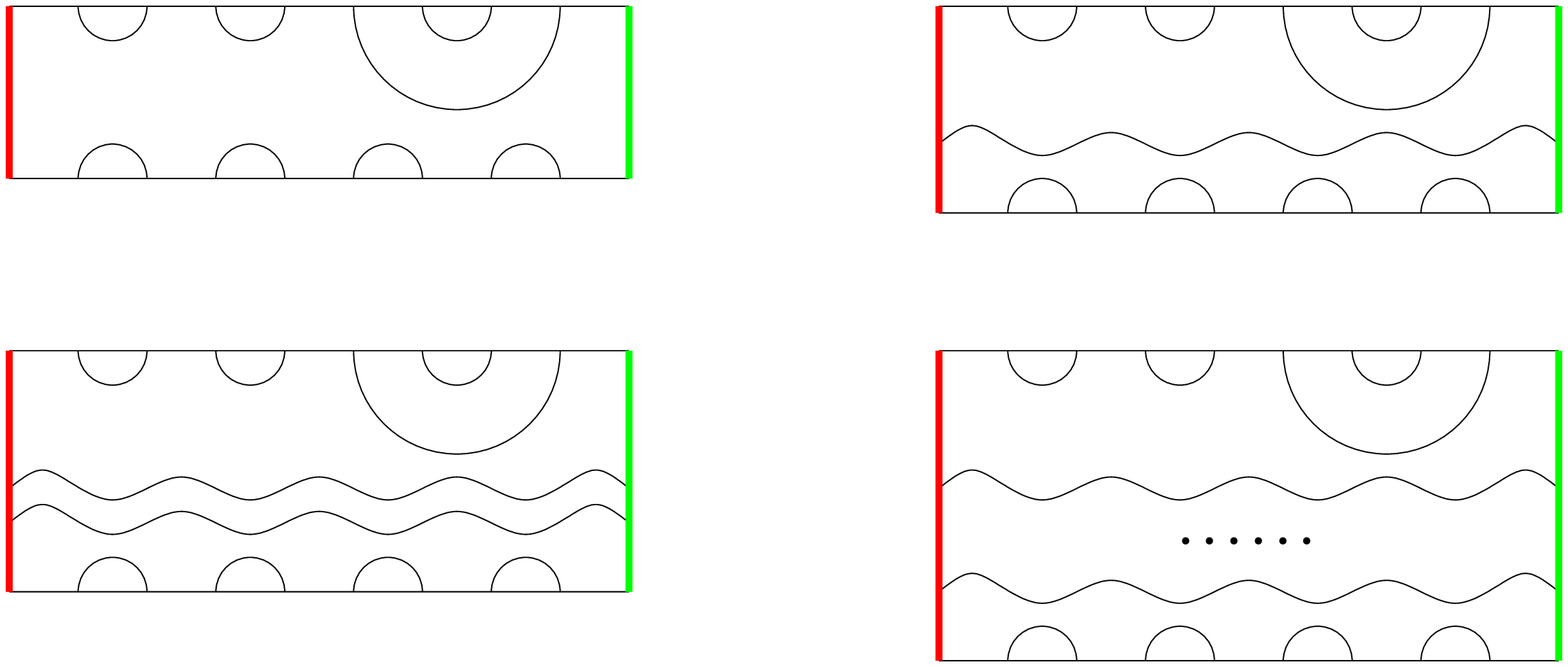}
\end{center}

However it has been proved \cite{degier_nichols} that all finite
dimensional irreducible 
representations come from a further the quotient of this algebra which
consist in giving a weight $\sqrt \tau_c$ to the lines going from one
boundary to the other. Algebraically one has to distinguish the case
$N$ odd and the case $N$ even, then introduce the following elements
\begin{itemize}
\item{\bf{$N = 2M+1$ :}}~~~~ $g_1 = f_L \prod_{i=1}^{M} e_{2i}$
  ~~~~$g_2=f_R\prod_{i=1}^{M} e_{2i-1}$ 
\item{\bf{$N = 2M$:}}~~~~  $g_1 = f_L f_R \prod_{i=1}^{M-1} e_{2i} $
  ~~~~$g_2=\prod_{i=1}^M e_{2i-1}$ 
\end{itemize}   
and take the quotients over
$$
g_1 g_2 g_1 =\tau_c g_1 ~~~~~ g_2 g_1 g_2 = \tau_c g_2.
$$
In the rest of the paper we will moreover restrict to the case
$\tau_R=\tau_L=1$. The reason will be explained in Appendix
\ref{recursion_bound}.

\subsection{Representation of $TL^{(o,o)}_N$ on extended link
  patterns}\label{extended }

The representation of $TL^{(o,o)}_N$ we will be interested in acts on the
space $\sH_N^{(oo)}$ with basis labelled by \emph{ extended link
  patterns}. This is the Hilbert space corresponding to open-open
boundary conditions. We call
``extended link pattern'' a diagram with 
$N$ points on a line, numbered from 
left to right, and two more points, the first called $L$ is
situated on the left of point $1$; the second called $R$ is situated on the
right of point $N$. The point $1, \dots N$ are either connected in pairs, or
they are connected to the point $L$ or $R$, by non intersecting curves. Here is an example
\sk
\begin{center}
\includegraphics[scale=.6]{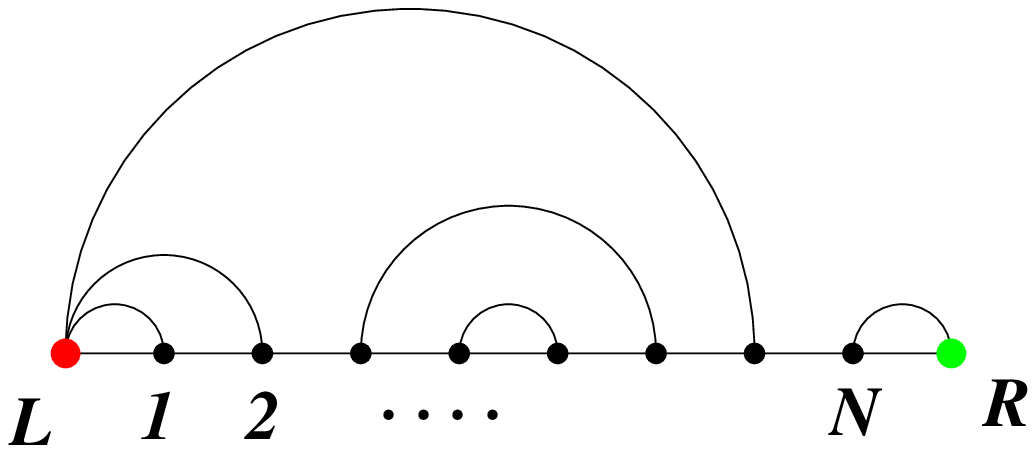}
\end{center}
The action of the generators  $TL^{(o,o)}_N$ is almost obvious from
the graphical representation, for simplicity we restrict to the
case $\tau_L=\tau_R=1$. In such a case each time we close a loop in
the bulk we remove it and multiply the link pattern by $\tau$. If
instead we close a loop touching one of the two boundaries we simply remove
it. 
\sk
\begin{center}
\includegraphics[scale=.6]{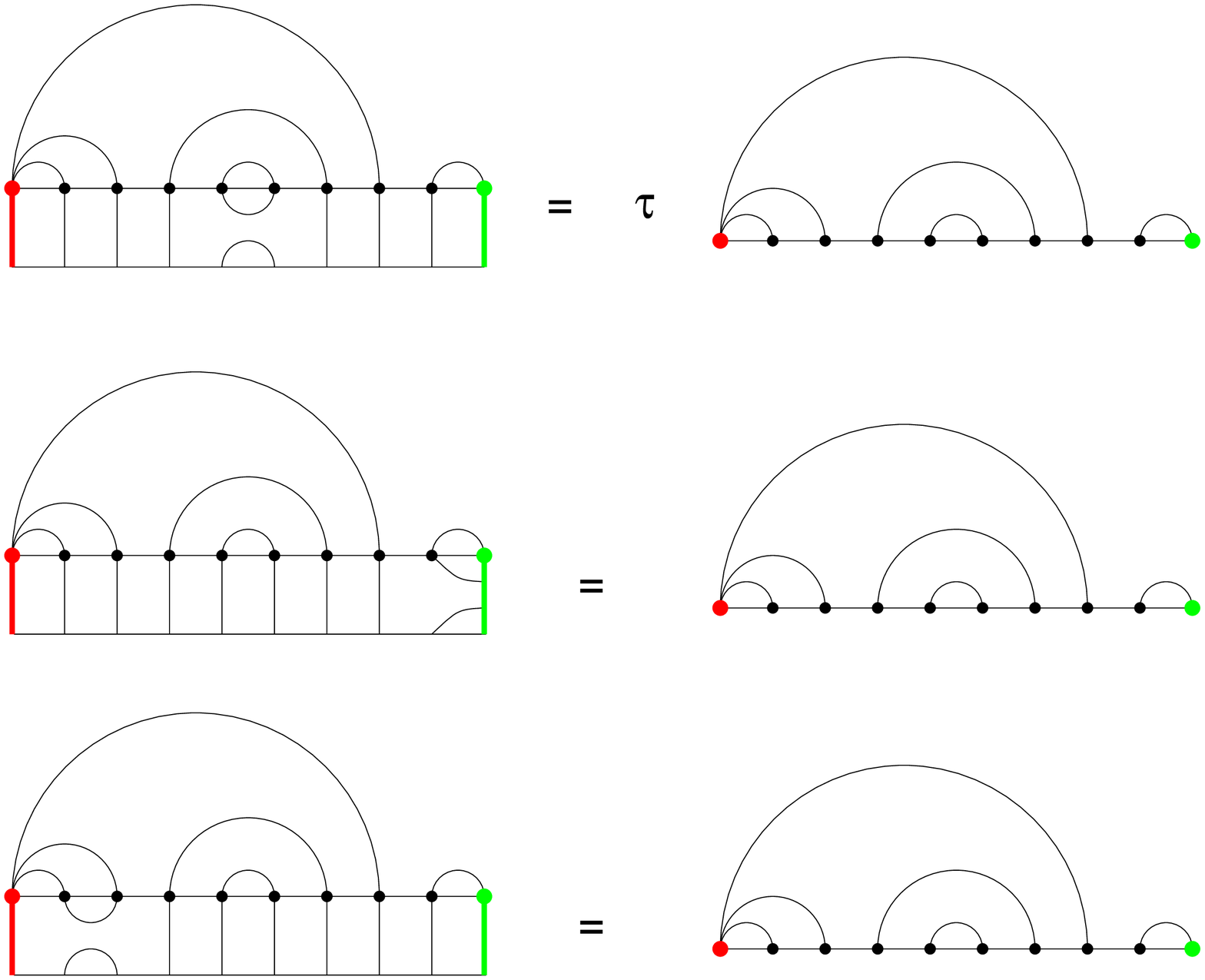}
\end{center}
\sk
These rules are supplemented by the requirement that a line
joining the two boundaries can be removed at the cost of multiplying
the obtained link pattern a weight. We call the weight of the removed
line $\sqrt{\tau_c}$. Graphically this looks as follows 
\sk
\begin{center}
\includegraphics[scale=.6]{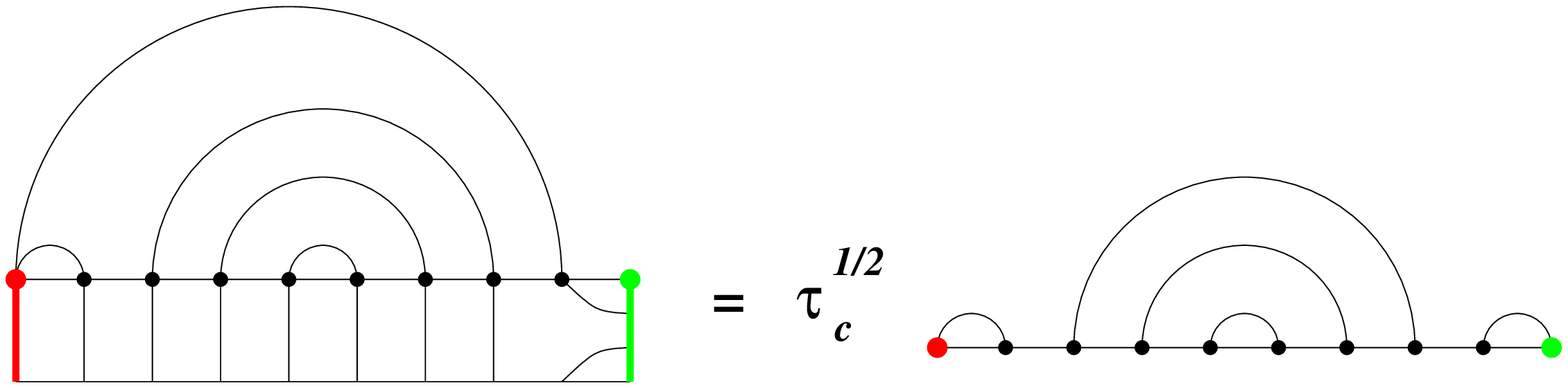}
\end{center}
\sk

In Section \ref{projection} we will consider also the representations of
$TL^{(o,c)}_N$ on the space of left extended link
patterns that we 
call $\sH_N^{(oc)}$. This space is the subspace of
$\sH_N^{(oo)}$ consisting of link patterns having no lines connected to
the point $R$ and is the Hilbert space for
open-closed b.c. considered in \cite{paul}.

\section{The boundary scattering matrix}\label{scatt_matr}

Let us consider now the following $\check R$-matrix
\vskip 1truecm
\eq\label{rcheck}
~~~~~~~= ~~\check{R}_i(z, w) = \frac{(qz -w/q)I + (z-w)e_i}{qw - z/q}
\en
\begin{picture}(0,0)
\put(85,-5){\includegraphics[scale=.4]{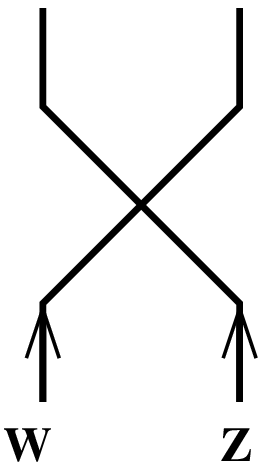}}
\end{picture}
\sk
\noindent
where $I$ is the identity operator and $\tau = -(q + q^{-1})$ (recall
that $e^2 = \tau e$).
The matrix $\check{R}_i$ satisfies the Yang-Baxter equation
$
\check{R}_{i+1}(w,z)\check{R}_{i}(x,z)\check{R}_{i+1}(x,w) =
\check{R}_{i}(x,w)\check{R}_{i+1}(x,z)\check{R}_{i}(w,z).  
$
\sk
\begin{center}
\includegraphics[scale=0.5]{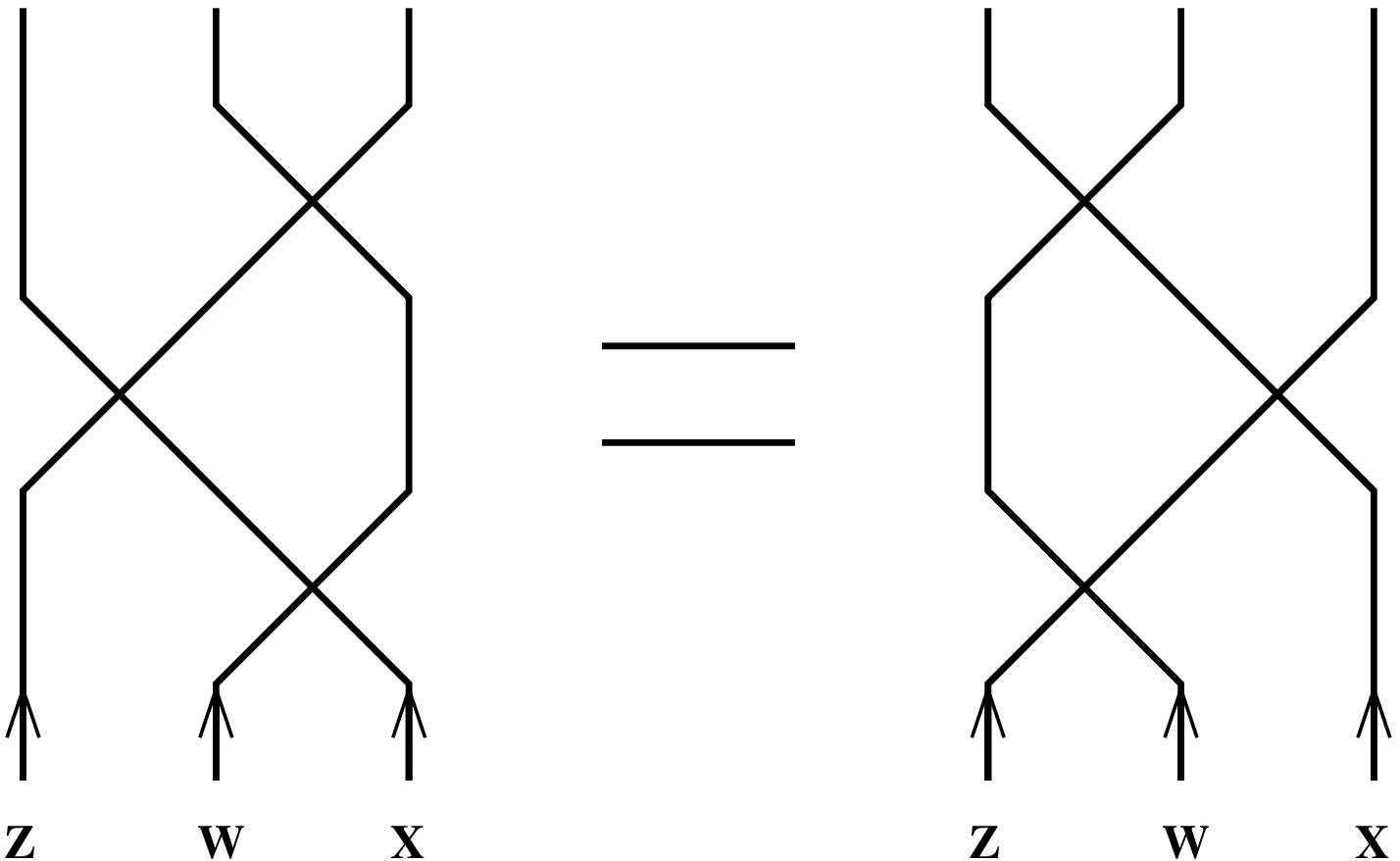}
\end{center}
\sk
\noindent
In a model with boundaries, integrability is assured by the presence
of boundary scattering matrix which satisfies the so called Boundary
Yang Baxter Equation \cite{sklyanin, cherednik_bound}
$$
K_R(w)\check{R}_{N-1}(1/z,w)K_R(z)\check{R}_{N-1}(w,z)=\check{R}_{N-1}(1/z,1/w)
K_R(z)\check{R}_{N-1}(1/w,z) K_R(w).    
$$
\sk
\begin{center}
\includegraphics[scale=0.5]{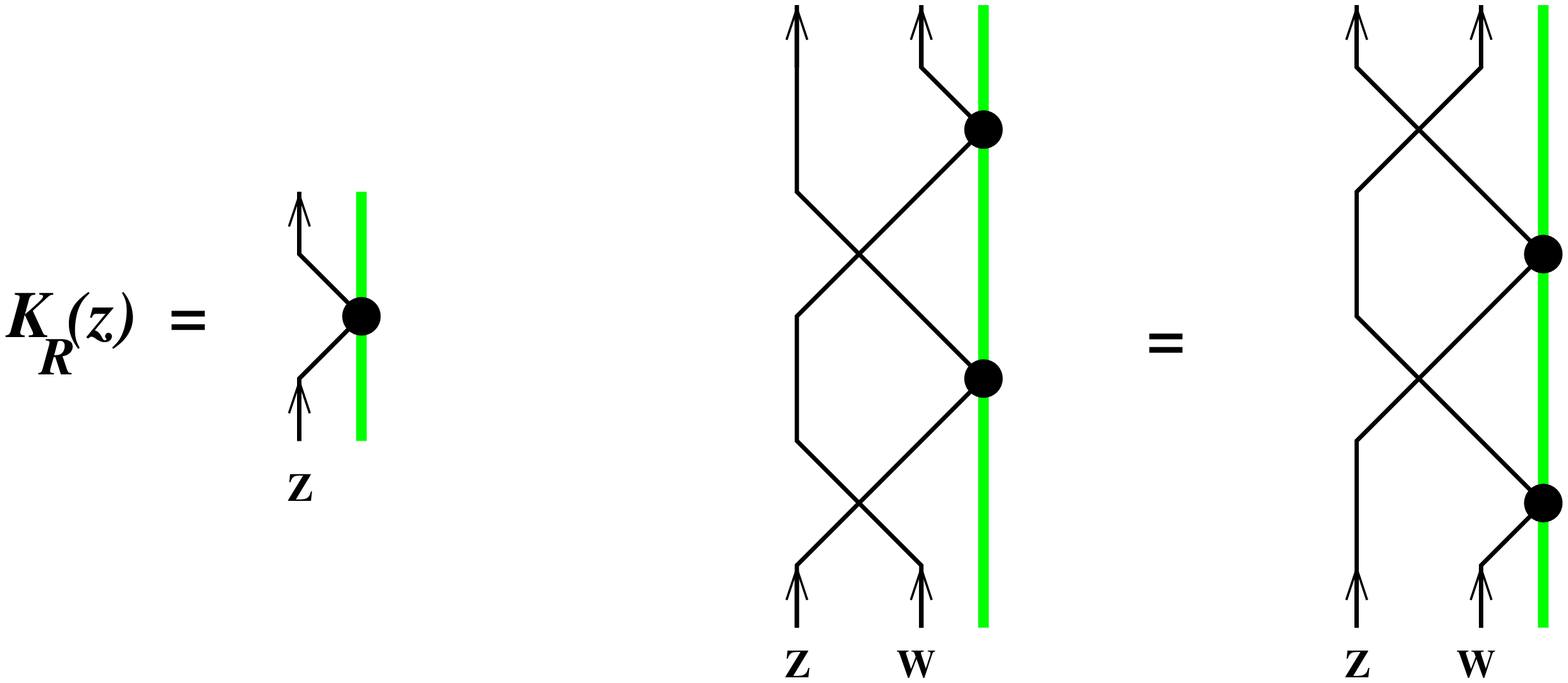}
\end{center}
\sk
In the present case the nontrivial boundary conditions on the right
are given by a right boundary 
scattering matrix $K_R$ of a Baxterized form $K_R(z) =
a(z)I + b(z)f_R$. The nontrivial solutions of the right boundary Yang Baxter
Equation  form a one parameter family, which in the case of generic
$\tau_R$ reads 
\eq\label{boundary_scatt}
K_R(\zeta_R|z) = \frac{(z-\zeta_R/q)(z-k q/\zeta_R)I +
  \frac{(q^2-1)}{(q+\tau_R)}(z^2-1)f_R}{(z-q/\zeta_R)(z k-\zeta_R/q)}.     
\en
with
$$
k=\frac{q+q^2\tau_R}{q+\tau_R} ~;~~~~ \tau_R =
q~\frac{k-1}{q^2 - k}.
$$
In the case $\tau_R=1$, which turns out to be the most interesting one
for us and to which we restrict from now on, we have $k=q$ and the
boundary scattering 
matrix reduces to 
\eq
K_R(\zeta_R|z) = \frac{(z-q^2/\zeta_R)(z- \zeta_R/q)I +
  (q-1)(z^2-1)f_R}{(qz-\zeta_R/q)(z-q/\zeta_R)}. 
\en
We consider left boundary scattering matrices as well but, in order
to formulate the qKZ equations, we require
them to satisfy a modified boundary Yang Baxter equation in which we
have a new parameter $s$
$$
\check{R}_{1}(s/w,s/z)K_L(z)\check{R}_{1}(z,s/w)K_L(w) = 
K_L(w) \check{R}_{1}(w,s/z) K_L(z) \check{R}_{1}(z,w).
$$
The explicit form of the left scattering matrices when $\tau_L=1$ is
\eq
K_L(\zeta_L|z) = \frac{(q z -s\zeta_L/q)(z-q/\zeta_L)+(q-1)(s-
  z^2)f_L}{(z -s\zeta_L/q)(z -q^2/\zeta_L)}.  
\en

\section{The qKZ equations and basic properties of their
  solutions}\label{qKZ-section} 

Given the representation of $TL^{(o,o)}_N$ on the space of extended
link patterns, let us consider a function 
$\Psi_N(\zeta_L;z_1,\dots, z_N;\zeta_R)$ from $\bC^{N+2}$ to this space.
The boundary qKZ equations are the following set
of equations for $\Psi(\zeta_L;z_1,\dots, z_N;\zeta_R)$
\eq\label{qKZ1}
\check{R}_{i}(z_{i+1},z_i)\circ \Psi(\zeta_L;z_1,\dots, z_i,
z_{i+1},\dots, z_N;\zeta_R) = 
\Psi(\zeta_L;z_1,\dots, z_{i+1}, z_{i},\dots, z_N;\zeta_R);
\en
\eq\label{qKZ2}
K_R(\zeta_R|z_N)\circ \Psi(\zeta_L;z_1,\dots, z_N;\zeta_R) =
\Psi(\zeta_L;z_1,\dots, 1/z_N;\zeta_R); 
\en
\eq\label{qKZ3}
K_L(\zeta_L|z_1)\circ \Psi(\zeta_L; z_1,\dots, z_N\zeta_R) =
\Psi(\zeta_L;s/z_1,\dots, z_N\zeta_R); 
\en
We expand now these equations on the natural basis of extended link
patterns. Therefore we write
$$
\Psi^{N}(\zeta_L;z_1,\dots, z_N;\zeta_R) = \sum_\pi
\psi^N_\pi(\zeta_L;z_1,\dots, z_N;\zeta_R) |\pi\rangle,
$$
where the sum runs over all the extended link patterns. 
Since given an extended link pattern, the notion of an arc opening or
closing at a point is well defined, we will parameterize it by the
location of open and closing arcs and use the
following notation $|coo\dots oc\rangle$,
each $o$ or $c$ standing for ``opening'' or ``closing''. A concrete
example of our notation is
\vskip 1truecm
$$
~~~~~~~~~~~~~~~~~~~~~~~~~~~~~~~~~~~~~~~~~~~~= ~~|ccooccco \rangle
$$
\begin{picture}(0,0)
\put(70,-5){\includegraphics[scale=.6]{exm-ext.eps}}
\end{picture}
\sk
\noindent
It is useful to introduce the following operators on the space of
Laurent polynomials of $N$ variables
\begin{equation}
t_i \circ \phi (\dots,z_i, z_{i+1},\dots) = \phi (\dots,z_{i+1},
z_{i},\dots),  
\end{equation} 
\begin{equation}
t_R \circ \phi (\dots,z_N) = \phi (\dots,1/z_N),
\end{equation} 
\begin{equation}
t_L \circ \phi (z_1,\dots) = \phi (s/z_1,\dots).
\end{equation} 
These operators allows us to rewrite the qKZ equations for the
components of $\Psi_N(\zeta_L;z_1,\dots, z_N;\zeta_R)$ in the extended
link patterns basis as follows. 
\begin{itemize}

\item From eq.(\ref{qKZ1}) it follows that if $|\pi > \notin e_i\circ \sH_{N}^{(oo)}$ then 
  we have
\begin{equation}\label{qKZ1comp1}
(q z_{i+1} - z_{i}/q)\psi^N_\pi = t_i \circ (q z_{i+1} - z_{i}/q)\psi^N_\pi
\end{equation}
Otherwise, if $|\pi > \in e_i \circ \sH_{N}^{(oo)}$ then
\begin{equation}\label{inv1}
\partial_i \psi^N_\pi :=(q z_{i} - z_{i+1}/q)\frac{1-t_i}{z_i-z_{i+1}} \psi^N_\pi= \sum_{\pi
  ~\propto ~e_i\circ \pi'} c^i_{\pi,\pi'} \psi^N_{\pi'},  
\end{equation}
where the coefficients $c^i_{\pi,\pi'}$ are defined by $e_i \pi' =c^i_{i,\pi,\pi'}
\pi $ and can be either $\sqrt{\tau_c}$ if in applying $e_i$ to
$\pi'$ we form a line joining the two boundaries or $1$ in case we do
not form a line joining the two boundaries. 
\item
From eq.(\ref{qKZ2}) it follows that if  $|\pi > \notin f_R\circ \sH_{N}^{(oo)}$ then 
\begin{equation}\label{qKZ2comp1}
(z_N-q^2/\zeta_R)(z_N- \zeta_R/q) \psi^N_\pi=(qz_N-\zeta_R/q)(z_N-q/\zeta_R)
t_R\circ \psi^N_\pi.
\end{equation}
If $|\pi > \in f_R \circ \sH_{N}^{(oo)}$ there is only one preimage under
$f_R$ different from $\pi$ itself,
i.e. there is only one $\pi'\neq \pi$ such that $\pi \propto f_R \circ
\pi'$ and the
components of this preimage is given by 
\begin{equation}
 c^R_{\pi,\pi'} \psi^N_{\pi'} = \partial_R \psi^N_\pi := (qz_N-\zeta_R/q)(z_N-q/\zeta_R)
  \frac{1-t_R}{(q-1)(z_N^2-1)}\circ \psi^N_\pi
\end{equation}
The coefficient $c^R_{\pi,\pi'}$ now is either $\sqrt{\tau_c}$ if by
applying $f_R$ to $\pi'$ we form a line joining the two boundaries,
or otherwise it is equal to $1$. 
\item
Finally from eq.(\ref{qKZ3}) it follows that if  $|\pi > \notin f_L
\circ \sH_{N}^{(oo)}$ then 
\begin{equation}
(q z_1 -s\zeta_L/q)(z_1-q/\zeta_L) \psi^N_\pi=(z_1-s\zeta_L/q)(z_1-q^2/\zeta_L)
t_L\circ \psi^N_\pi.
\end{equation}
If $|\pi > \in f_L \circ \sH_{N}^{(oo)}$ its preimage under $f_L$, different
from $\pi$ itself is unique
\begin{equation}
 c^L_{\pi,\pi'} \psi^N_{\pi'} = \partial_L \psi^N_\pi := (z_1-s\zeta_L/q)(z_1-q^2/\zeta_L)
  \frac{1-t_L}{(q-1)(s-z_N^2)}\circ \psi^N_\pi
\end{equation}
The coefficient $c^L_{\pi,\pi'}$ now is either $\sqrt{\tau_c}$ if by
applying $f_L$ to $\pi'$ we form a line joining the two boundaries,
or otherwise it is equal to $1$.
\end{itemize}

\subsection{Affine Hecke generators}\label{affine_sectio} 

It is known, from the seminal papers \cite{cherednik_qKZ, kato}, that the
qKZ equations are related to the representation theory of affine Hecke
algebras. This has been rediscovered recently in the context of the
Razumov Stroganov conjecture \cite{pasquier, pasquier_kasatani}. 

Let us
explain this observation in our case; this will lead us to consider the 
Laurent polynomial representations of affine Hecke algebras of type C
introduced by Noumi \cite{noumi}. We start from eq.(\ref{qKZ1}), and
introduce different generators of the $TL_N^{(o,o)}$ algebra $T_i=
-e_i-1/q$. By recombining the terms of eq.(\ref{qKZ1}) we can rewrite
it as
\begin{equation}
T_i\circ  \Psi_N = \hat T_i \circ \Psi_N,
\end{equation}
where the operator 
$$
\hat T_i = q + \frac{1}{q}\left(\frac{q^2z_i
  -z_{i+1}}{z_i-z_{i+1}}\right)(t_i-1) 
$$
acts on the polynomial part of $\Psi_N$. We proceed in the same way
for the other two qKZ equation, by introducing two generators 
$T_N= (q_N+1/q_N)f_R-1/q_N$ and $T_0= (q_0+1/q_0)f_L-1/q_0$, where
$q_0^2 = q_N^2=-q$. Then we can rewrite eqs.(\ref{qKZ2},\ref{qKZ3}) as
\begin{equation}
T_0\circ \Psi_N = \hat T_0\circ \Psi_N,~~~~~T_N\circ \Psi_N = \hat
T_N\circ \Psi_N. 
\end{equation}
The operators $\hat T_0$ and $\hat T_N$, like the $\hat T_i$, act on
Laurent Polynomials and are given by 
$$
\hat T_0 = q_0 + \frac{1}{q_0} \frac{\left(z-\frac{s\zeta_L}{q}
  \right) \left(z -\frac{q^2}{\zeta_L} \right)}{(z^2-s)}(t_L-1).
$$
$$
\hat T_N = q_N +
\frac{1}{q_N}\frac{\left(1-\frac{q^2z}{\zeta_R}\right)
  \left(1-\frac{z\zeta_R}{q} \right)}{(1-z^2)}(t_R-1)
$$
It is now a matter of some straightforward computations to show that
the operators $\{T_i,T_0,T_N\}$ and  $\{\hat T_i,\hat T_0,\hat T_N \}$ 
satisfy separately the 
commutation relations of the generators of the affine Hecke algebra
$\mathcal{H}(C_N)$ of type $C_N$
\begin{equation}
\begin{array}{l}
(T_0+1/q_0)(T_0-q_0)=0\\(T_i+1/q)(T_i-q)=0\\(T_N+1/q_N)(T_N-q_N)=0\\
T_iT_{i\pm 1}T_i= T_{i\pm 1}T_iT_{i\pm
  1}\\T_0T_1T_0T_1=T_1T_0T_1T_0\\T_NT_{N-1}T_NT_{N-1}=T_{N-1}T_NT_{N-1}T_N\\
T_iT_j=T_jT_i~~~\textrm{for} ~~~|i-j|>1\\
T_0T_j=T_jT_0~~~\textrm{for} ~~~j>1\\
T_NT_j=T_jT_N~~~\textrm{for} ~~~j<N-1\\
\end{array}
\end{equation}
Indeed the representation of $\mathcal{H}(C_N)$ given by $\{\hat
T_i,\hat T_0,\hat T_N \}$ is well 
known and goes under the name of Noumi representation (actually in the
Noumi representation the parameters $q_0$, $q_N$ and $q$ are
independent; see for example Proposition 2.2 of \cite{kasatani}). By
adding also the operators $\hat z_i$, whose action on a polynomial is
the multiplication by $z_i$ we obtain a representation of the doubly
affine Hecke algebra of type $C^{\vee}C_N$ \cite{sahi, kasatani}.

In the qKZ equation, we are considering the action of two copies of the
$\mathcal{H}(C_N)$, one acting on $\sH_N^{(oo)}$, the other acting on
a space of polynomial that for the moment we call $\hat
\sH_N^{(oo)}$. Then the vector $\Psi_N$ can be interpreted as a map
from $\sH_N^{(oo)}$ to the dual of $\hat\sH_N^{(oo)}$
$$\Psi_N: \sH_N^{(oo)} \rightarrow \hat \sH_N^{(oo)*}$$ and the qKZ
equations simply state that it intertwines between the two
representation. Therefore, since the representation on $\sH_N^{(oo)}$
is irreducible \cite{degier_nichols}, we conclude that $\hat
\sH_N^{(oo)} = \sH_N^{(oo)*}$. This means that solving the qKZ
equations amounts first to find the irreducible representation of
$\mathcal{H}(C_N)$ on Laurent polynomials, dual to $\sH_N^{(oo)}$, and
then to find the basis dual to the extended link pattern basis of
$\sH_N^{(oo)}$.

\subsection{Trivial factors and symmetries}\label{trivial_factors}

From eq.(\ref{qKZ1comp1}) it follows that a component in which the points
$i$ and $i+1$ are not connected, will have the following form
$$
\psi(z_i,z_{i+1}) = (q z_i - z_{i+1}/q) \tilde \psi(z_i,z_{i+1})
$$
where $\tilde \psi(z_i,z_{i+1})$ is symmetric under exchange $z_i
\leftrightarrow z_{i+1}$. In general if the consecutives  points $i,i+1,\dots,
i+r$ are not connected among themselves then
$$
\psi(z_i,z_{i+1},\dots ,z_{i+r}) = \prod_{i\leq j <k \leq i+r}(q z_j -
z_{k}/q) ~\tilde \psi(z_i,z_{i+1},\dots ,z_{i+r}) 
$$
and $\tilde \psi(z_i,z_{i+1},\dots ,z_{i+r})$ is symmetric in
$z_i,\dots,z_{i+r}$.  

Analogously, from eq.(\ref{qKZ2comp1}), it follows that if the point $N$ is not
connected to the right boundary $R$ then one has
$$
\psi(z_N;\zeta_R)=(q z_N/\zeta_R-1/q)(\zeta_R-q/z_N)
\tilde\psi(z_N;\zeta_R ) 
$$
where $\tilde\psi(z_N;\zeta_R )$ is invariant under $z_N\rightarrow
1/z_N$.
An analogous statement relative to the left boundary is true, namely
if the point $1$ is not connected to the left boundary $L$ then the
component can be written as
$$
\psi(\zeta_L;z_1) = (z_1/\zeta_L-s/q)(\zeta_L-q^2/z_1)\tilde \psi(\zeta_L;z_1)
$$
where now $\tilde \psi(\zeta_L;z_1)$ is invariant under
$z_1\rightarrow s/z_1$.

One can combine the previous remarks in order to extract more trivial
factors. For example the
components having all the bulk points connected to the right boundary
$\psi_{oo\dots oo} $ or to the left boundary $\psi_{cc\dots cc}$ have
the following form 
\begin{equation}\label{factor-right}
\psi\underbrace{_{oo\dots oo}}_N(\zeta_L;\vec z;\zeta_R) =
\prod_{j=1}^N\left(\frac{s q
    \zeta_L}{z_j}-q^2\right)\left(\frac{z_j}{q}-\frac{q}{\zeta_L}\right)
\prod_{1\leq i<j \leq N}
(\frac{q  
  z_i}{z_{j}} - \frac{1}{q})(\frac{z_j}{q} -\frac{q s}{z_i } )
\phi^{(R)}_N(\zeta_L;\vec z;\zeta_R)
\end{equation}
\begin{equation}\label{factor-left}
\psi\underbrace{_{cc\dots cc}}_N(\zeta_L;\vec z;\zeta_R) =
\prod_{i=1}^N\left(\frac{q z_i}{\zeta_R}-\frac{1}{q}\right)\left(
  q\zeta_R-\frac{q^2}{z_i}\right) 
\prod_{1\leq i<j \leq N} (\frac{q 
  z_i}{z_{j}} - \frac{1}{q})(q z_j  - \frac{1 }{q z_i})
\phi^{(L)}_N(\zeta_L;\vec z;\zeta_R),
\end{equation}
where $\phi^{(R)}_N(\zeta_L;\vec z;\zeta_R)$ is a function
symmetric under exchange $z_i\leftrightarrow z_j$ and invariant under
$z_j\rightarrow s/z_j$, while $\phi^{(L)}_N(\zeta_L;\vec z;\zeta_R)$
is a function symmetric under exchange $z_i\leftrightarrow z_j$ and
invariant under $z_j\rightarrow 1/z_j$. 


\subsection{Recursion relations: bulk}\label{app_rec_bulk}

As a consequence of the analysis in the previous section, it follows 
that if we set $z_{i+1}= q^2 z_i$, all the components of the solution
of the qKZ equation at length $N$ that do not lie in the image of
$e_i$ are zero.
The subspace of $TL^{(o,o)}_N$ with an arc between
points $i$ and $i+1$ is 
isomorphic to the space $TL^{(o,o)}_{N-2}$. More precisely,  if we call $p_i$
the map $p_i: 
TL^{(o,o)}_{N-2}\rightarrow TL^{(o,o)}_N$ which consists in adding an arc between
$i-1$ and $i$ (and renumbering the points), then $p_i$ is an
isomorphism between $TL^{(o,o)}_{N-2}$ and $e_i \circ TL^{(o,o)}_{N}$.
Since $\Psi_N(\dots, z_i, z_{i+1}=q^2 z_{i},\dots)\in e_i \circ TL^{(o,o)}_{N}$ we
can consider its preimage under $p_i$. We show now that such a
preimage, $p_i^{-1}\circ\Psi_N(\dots, z_i, z_i=q^2 z_{i},\dots)$,
satisfies a set of
modified qKZ equations. In order to work this out let us restrict for
the moment to the case $i=N-1$, i.e. when we add a small arc at right the
end of our strip.  
It is easy to show that the map $p_{N-1}$ intertwines the operators
$K_L$ and $\check{R}_{i< N-2}$ acting on $TL^{(o,o)}_{N-2}$ and the ones acting on
$e_i \circ TL^{(o,o)}_{N}$ (by abuse of notation we do not adopt a different
notation for operators acting on different spaces)
$$
K_L(\zeta_L| z)~p_{N-1}=p_{N-1}~ K_L(\zeta_L| z),~~~~~
\check{R}_{i<N-2}(z,w)~ p_{N-1}=p_{N-1}~ \check{R}_{i<N-2}(z,w).
$$
Then let us consider the following matrix 
\begin{equation}
\bar K_R(\zeta_R|z)=\check R_{N-2}(1/z,q^2 z_{n-1})\check R_{N-1}(1/z,q^2
z_{n-1})K_R(\zeta_R|z)\check R_{N-1}(q^2 z_{n-1},z)\check R_{N-2}(z_{n-1},z). 
\end{equation}
When this matrix acts on configurations in the image of $e_{N-1}$, it
can be written as  
\begin{equation}
\bar K_R(\zeta_R|z)~ e_{N-1} = \frac{u(1/z)}{u(z)}
\left(\frac{(z-\zeta_R/q)(z-q^2/\zeta_R)I + 
  (q-1)(z^2-1)e_{N-1}f_R e_{N-2}}{(z\zeta_R/q-1)(z
  q^2/\zeta_R-1)}  \right)~ e_{N-1},
\end{equation}
where $u(z) = (q^4 z_{N-1}-1/z)(q^2 z -z_{N-1})$. Moreover if we
notice that $e_{N-1}f_R e_{N-2} p_{N-1} = p_{N-1} f_R$
\sk
\begin{center}
\includegraphics[scale=.6]{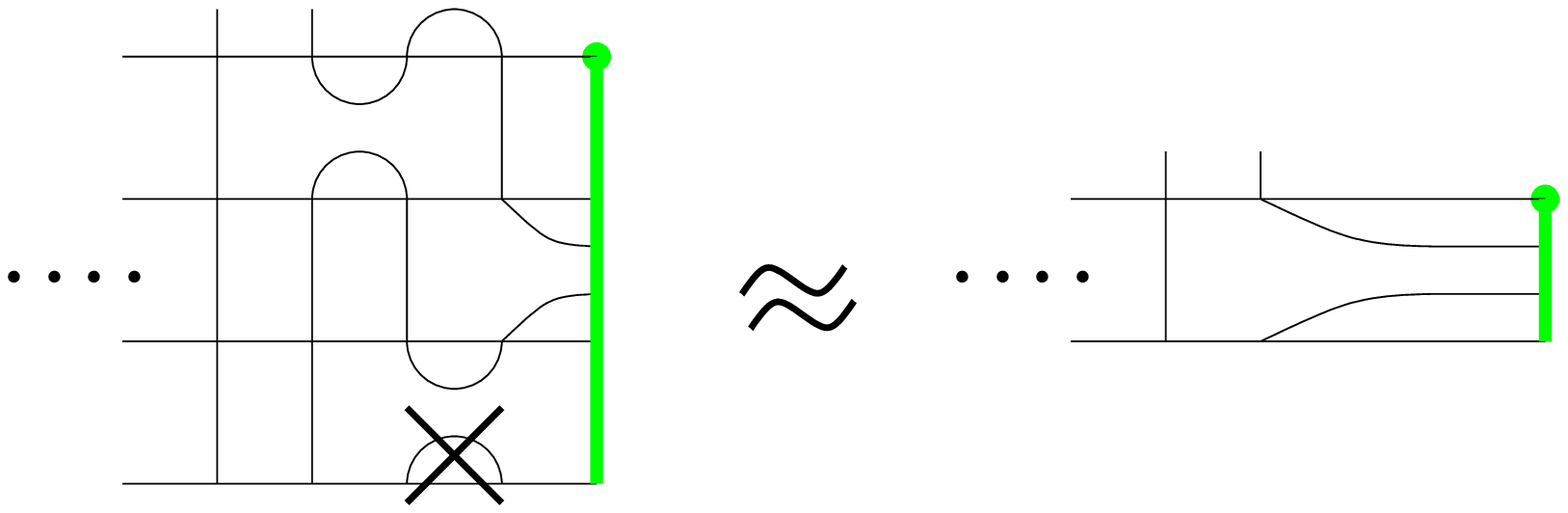}
\end{center}
\sk
we obtain that $p_{N-1}$ intertwines between $\bar K_R(\zeta_R|z)$ and
$u(1/z)/u(z) K_R(\zeta_R|z)$
\begin{equation}
\bar K_R(\zeta_R|z)~ p_{N-1} =
\frac{u\left(1/z\right)}{u(z)}p_{N-1} ~ K_R(\zeta_R|z).
\end{equation}
Therefore $p^{-1}_i\circ
\Psi_N(\dots, z_{N-1},z_N= q^2 z_{N-1},\dots) $ satisfies a 
modified set of qKZ equations where eq.(\ref{qKZ2}) is substituted by
the following equation
\eq\label{qKZ2modif}
\begin{array}{c}
\frac{u\left(1/z_{N-2}\right)}{u(z_{N-2})}K_R(\zeta_R|z_{N-2})
p^{-1}_{N-1}\circ \Psi_N(\dots, z_{N-2},z_{N-1}, z_N=q^2 z_{N-1},\dots)\\ 
=
p^{-1}_{N-1}\circ \Psi_N(\dots, 1/z_{N-2},z_{N-1}, z_N=q^2 z_{N-1},\dots)  
\end{array}
\en
Now let us suppose that
we can find a function $g(z)$ which is invariant under $z\rightarrow
1/z$ and such that
\begin{equation}\label{rel_rec}
\frac{u(z)g(z)}{u\left(\frac{s}{z}\right)g\left(\frac{s}{z}\right)}=1  
\end{equation}
then we have that 
$$
\prod_{i=1}^{N-2} \frac{1}{u(z_i)g(z_i)}~p^{-1}_{N-2}\circ
\Psi_N(\dots, z_{N-1}, z_N=q^2 z_{N-1},\dots) 
$$
is a solution of the unmodified qKZ equations
(\ref{qKZ1},\ref{qKZ2},\ref{qKZ3}) 
of length $N-2$ and. Therefore, apart from a factor independent from all
the $z_{i<N-1}$, it must be equal to $\Psi_{N-2}(\dots, z_{N-2};\zeta_R) $
\begin{equation}
 \Psi_N(\dots, z_{N-1}, q^2 z_{N-1},\dots)  =
k(z_{N-1},\zeta_L,\zeta_R ) \left( \prod_{i=1}^{N-2} u(z_i)g(z_i)\right)
p_{N-1}\circ\Psi_{N-2}(\dots, z_{N-2};\zeta_R)  
\end{equation}

Since we are searching for Laurent polynomial solutions of the qKZ we
require the function $g(z)$, solution of eq.(\ref{rel_rec}), to be also
a Laurent polynomial. This fixes the possible form of $s$ as function
of $q$. Indeed, in order to cancel the pole at $z= q^2 s/z_{N-1}$ in
eq.(\ref{rel_rec}), the 
function $g(z)$ must be of the form
\begin{equation}
g(z)=\prod_{j=1}^{n-1} \left(1-\frac{q^2 s^j}{z_{N-1} z}  \right)\left(
s^j z- z_{N-1}/q^2\right). 
\end{equation} 
The product must be such that the last term cancels the pole at $z=s
q^4 z_{N-1}$.  This happens only if $s=q^{-6/n}$.  

In Section \ref{sol_section} we will argue that the lowest value of
$n$ for which it is possible to find a solution of the qKZ equations is
$n=4$ or $s=q^{-3/2}$. In such a case 
the recursion relation takes the form
\begin{equation}
\Psi_N(\dots, z_{N-1}, z_N= q^2 z_{N-1},\dots)=
G(z_1,\dots;z_{N-1},\zeta_L,\zeta_R)~p_{N-1}\circ \Psi_{N-2}(\dots, \hat   
z_{N-1}, \hat z_{N},\dots).
\end{equation}
with
\begin{equation}
G(z_1,\dots;z_{N-1},\zeta_L,\zeta_R):=k(z_{N-1},\zeta_L,\zeta_R )\prod_{i=1}^{N-1}
\prod_{j=1}^4\left(q^{3j/2}-\frac{q^2 s^j}{z_{N-1} z_i}
\right)\left(z_i- q^{3(j-1)/2-2}z_{N-1}\right) 
\end{equation}
This is the recursion relation we were searching for. It will be very
important in the next section when we will construct a solution of the
qKZ equations. Actually we will need as well an analogous formula when
we set $z_{2}= q^2 z_1$ for $i=1$. Of course we could repeat almost
word by word the same derivation using this time a modified left
boundary scattering matrix. However, in doing so we would not be able
to keep track of the normalization choice of the wave function $\Psi_N$,
which is implicit in the function $k(z_{N-1},\zeta_L,\zeta_R )$. We
can avoid this problem by deriving the recursion relation when
$z_{i+1}= q^2 z_i$ from the case $i=N-1$.  The idea is simply to use
eq.(\ref{qKZ1}) to move the spectral lines. If we act on
$\Psi_N(\dots, z_{N-1}, z_N= q^2 z_{N-1},\dots)$ with the operator
$\check{R}_{N-1}(z_{N-1},z_{N-2})\check{R}_{N-2}(z_N=q^2z_{N-1}, z_{N-2})$ we
obtain  
\sk
\begin{center}
\includegraphics[scale=.6]{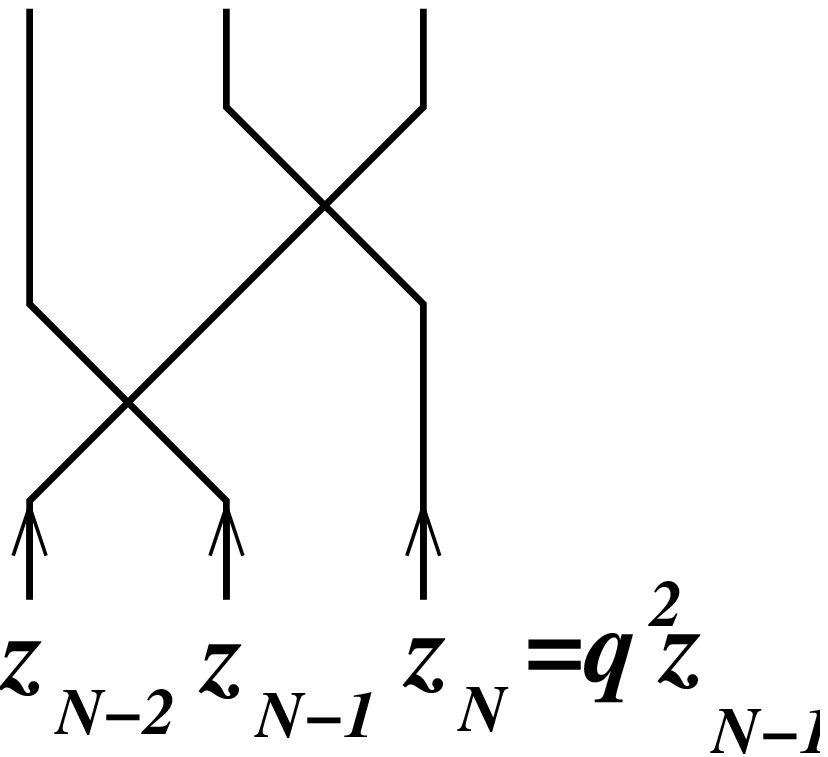}
\end{center}
\sk
$$
\Psi_N(\dots, z_{N-1}, z_N= q^2 z_{N-1},z_{N-2};\zeta_R)=
$$
$$
=\check{R}_{N-1}(z_{N-1},z_{N-2})\check{R}_{N-2}(z_N=q^2 z_{N-1}, z_{N-2})
\Psi_N(\dots,z_{N-2}, z_{N-1}, z_N = q^2 z_{N-1};\zeta_R) 
$$
$$
=G(z_1,\dots;z_{N-1},\zeta_L,\zeta_R)
\check{R}_{N-1}(z_{N-1},z_{N-2})\check{R}_{N-2}(z_N=q^2  
z_{N-1}, z_{N-2}) p_{N-1}\Psi_{N-2}(\dots,z_{N-2};\zeta_R), 
$$
where the second equality follows from the recursion relation proved
above. 
Then we notice that when we act with
$\check{R}_{N-1}(z_{N-1},z_{N-2})\check{R}_{N-2}(z_N=q^2, z_N-2)$ on
$e_{N-1} TL^{(o,o)}_N$ we have
\begin{equation}\label{transport1}
 \check{R}_{N-1}(z_{N-1},z_{N-2})\check{R}_{N-2}(z_N=q^2, z_{N-2})
 e_{N-1} = f(z_{N-1}, z_{N-2}) e_{N-2}e_{N-1},
\end{equation}
with $$f(z_{N-1}, z_{N-2})=\frac{q z_{N-1}-q^3 z_{N-2}}{q^4 z_{N-1}-z_{N-2}}.$$
This property can be restated in terms of
intertwinings, just by substituting $e_{N-1}$ with $p_{N-1}$ in the
l.h.s. and  $e_{N-2}e_{N-1}$ with $p_{N-2}$ in the
r.h.s. of eq.(\ref{transport1}). Therefore we arrive at
\begin{equation}
\begin{array}{c}
\Psi_N(\dots, z_{N-1}, z_N= q^2 z_{N-1},z_{N-2};\zeta_R) \\
=
G(z_1,\dots;z_{N-1},\zeta_L,\zeta_R)
f(z_{N-1},z_{N-2})p_{N-2}\Psi_{N-2}(\dots,z_{N-2};\zeta_R) 
\end{array} 
\end{equation}
We can continue this way and move the lines $N-1$ and $N$ to the
position $i$ and $i+1$. By shuffling the indices we obtain therefore
\begin{equation}
\Psi_N(\dots, z_{i}, z_{i+1}= q^2 z_{i},\dots) =
G(\dots)\prod_{j>i+1}f(z_i,z_j)~p_{i}\circ\Psi_{N-2}(\dots,\hat z_i, \hat
z_{i+1},\dots),  
\end{equation}
where the hats in the right hand side mean that the variables are
removed. 

\section{Solution at $s=q^{-3/2}$}\label{sol_section}

We search for a solution of the qKZ equations
(\ref{qKZ1},\ref{qKZ2},\ref{qKZ3}), in the simplest situation,
i.e. when we have only two internal points ($N=2$). 
For the moment we leave the
parameters $s$ and $\tau_c$ free and we will see whether they will be
fixed by the solution.
Let us start from the components $\psi_{oo}(\zeta_L; z_1,z_2;\zeta_R)$
and $\psi_{cc}(\zeta_L; z_1,z_2;\zeta_R)$. We have seen in Section
\ref{trivial_factors} that they have the form reported in equation
eqs.(\ref{factor-right}, \ref{factor-left}). Some tedious but
straightforward calculations show that if we demand the Laurent
polynomials $\chi^{(R)}_2(\zeta_L;z_1, z_2;\zeta_R)$ and
$\chi^{(L)}_2(\zeta_L;z_1, z_2;\zeta_R)$  to be constant in the $z_i$, then
there are no solutions of the qKZ equation. If instead we allow the
$\chi$s to be of degree width $1$, we see that the requirement that 
\begin{equation}
\psi_{oc} = \partial_R \psi_{oo} = \partial_L \psi_{cc}
\end{equation}
determines $s=q^{-3/2}$ and fixes completely (except of course for an
irrelevant global constant normalization) both $\phi^{(R)}_2$ and
$\phi^{(L)}_2$ 
\begin{equation}
\phi^{(R)}_2= \left(z_1 + \frac{1}{q^{3/2}z_1}+ z_2  +
\frac{1}{q^{3/2}z_2} \right) - (1+q^{-1/2})\left(\frac{q}{\zeta_R}+\frac{\zeta_R}{q^2}
\right),
\end{equation}
\begin{equation}
\phi^{(L)}_2=  \left(z_1 +\frac{1}{z_1}+z_2 +\frac{1}{z_2}
\right)-
(1+q^{1/2})\left(\frac{q^2}{\zeta_L}+\frac{\zeta_L}{q^{5/2}}
\right). 
\end{equation}
We find $\psi_{co}$ by  applying $\partial_1$ to $\psi_{oc}$
$$
\psi_{co} = \frac{1}{\sqrt{\tau_c}} (\partial_1 \psi_{oc}-\psi_{oo}-
\psi_{cc}), 
$$
then the qKZ equations close if we have
$$
\partial_R \psi_{co} = \sqrt{\tau_c} \psi_{cc} ~~~~\textrm{ and}
~~~~\partial_L \psi_{co} = \sqrt{\tau_c} \psi_{oo}. 
$$
These conditions are satisfied only if $\tau_c=\frac{1}{q^{1/2}+2+q^{-1/2}}$.
Armed with this simple solution we try to construct the solution  at
$N=3$. We notice that the
representation of $TL^{(o,o)}_N$ on extended link patterns at
$\tau=-q-1/q$ and $\tau_c= \frac{1}{q^{1/2}+2+q^{-1/2}}$ is
irreducible, therefore the discussion in Section \ref{affine_sectio}
tells us that if 
we know a component of $\Psi_{N}(\zeta_L;\vec z; \zeta_R)$,
then using the qKZ equations as in \cite{pdf-pzj-1} we can reconstruct
all the 
other components. It turns out that making some minimal assumptions we are able to
construct $\phi^{(R)}_3$. For this we have to come back to the bulk 
recursion relations of Section \ref{app_rec_bulk}. In order for them to be really
useful we need to determine as much as possible the unknown factor
$k(z,\zeta_L,\zeta_R )$. First of all using the qKZ equation
(\ref{qKZ1}) we have
$$
\psi_{o\dots oc} = \partial_R \psi_{o\dots oo}
$$
moreover if we specialize at $z_N=q^2 z_{N-1}$ we have
$$
\psi_{o\dots oc}(z_N=q^2z_{N-1}) = \frac{\left(q^3
    z_{N-1}-\frac{\zeta_R}{q} \right)\left(q^2
    z_{N-1}-\frac{q}{\zeta_R} \right)}{(1-q)(q^4z^2_{N-1}-1)} \psi_{o\dots
  oo}(\dots,z_{N-1},1/(q^2z_{N-1})).
$$
Applying the bulk recursion relation to the left hand side and
comparing it to the right hand side (where we use 
eq.(\ref{factor-right})) we find a factor of $k(z_{N-1}, \zeta_L,\zeta_R)$
$$
k(z, \zeta_L,\zeta_R) \propto \left(\frac{sq\zeta_L}{z}
  -q^2 \right)\left(\frac{z}{q}-\frac{q}{\zeta_L}
\right)\left(s q^3 z-\frac{q^2}{\zeta_L}
\right)\left(\frac{\zeta_L}{q^3 z}-q \right)
$$  
Repeating the same argument using $\psi_{c\dots cc}$
and the recursion relation for $z_2=q^2 z_1$ we obtain a further
factor 
$$
k(z, \zeta_L,\zeta_R) \propto \left(\frac{q^2 s}{z\zeta_R}
  -1 \right)\left(\zeta_R-\frac{qz}{s}
\right)\left(\frac{q z}{\zeta_R}-\frac{1}{q^3}
\right)\left(\zeta_R -\frac{1}{qz} \right)
$$
Then we assume that $k(z,\zeta_L,\zeta_R )$ is only given by the
product of the two factors above.
With this expression for $k(z,\zeta_L,\zeta_R )$, the recursion
relation for $\phi_N^{(R)}$ reads 
\begin{equation}\label{ref_phi}
\phi^{(R)}_N(\zeta_L; z_j= sq^2z_{i}; \zeta_R) = \tilde
G(z_1,\dots;z_{N-1};\zeta_R) \phi^{(R)}_N(\zeta_L; \hat z_j,\hat z_{i}; \zeta_R)
\end{equation}
where
$$
\tilde G(z_1,\dots;z_{N-1};\zeta_R) = \frac{q^6(1-q)}{s(1-sq^4)}\left(\frac{q^2
    s}{z_i \zeta_R-1}  \right)\left(\zeta_R-\frac{q z_i}{s}
\right)\prod_{k\neq i,j}
\prod_{l=2}^3\left(s^{-l}-\frac{q^2}{z_i z_k}
\right)\left(z_k- \frac{s^{1-l}z_i}{q^2}\right) 
$$
The further assumption that we make is that these recursion relations
completely fix $\phi_N^{(R)}$. In Section \ref{reconstruction} we will
show how to 
construct the solution of our qKZ equation in the special case
$q^{1/2}=e^{-2\pi i/3}$, using the known solution of the problem with
mixed boundary conditions. The solution we will found has a degree
width $4N-2$ in each variables $z_i$, which means that the degree
width of $\phi_N^{(R)}$ in each variables $z_i$ is $2N-1$. If we
suppose that the same remains 
true for generic $q$ at $s=q^{-3/2}$ then the bulk recursion relations
allow to fix $\phi_N^{(R)}$ completely. Indeed thanks to its
symmetries, once given 
eq.(\ref{ref_phi}), we know the values of $\phi_N^{(R)}$ for $z_i= q^{\pm
  2} z_{j\neq i}^{\pm 1}$. We have constructed $\phi_N^{(R)}$, using
Lagrange interpolation in $z_N$, up to $N=6$. 
It is actually quite remarkable that the interpolating
formula turns out to be a Laurent polynomial in the
other variables 
even though it doesn't look of this form. This is also the case
for all its expected symmetries.
Once $\psi\underbrace{_{oo\dots
    oo}}_N$ known, we have checked up to length $N=6$ that the qKZ
equations 
hold. We believe that some representation theoretical argument should
prove that the construction is valid for generic $N$.

\section{The solution at $q^{1/2}=e^{-2\pi i/3}$ : norm and
  homogeneous limit}\label{specialization}

As mentioned in the introduction, when the parameter $q^{1/2}$ assumes
the value $e^{-2\pi i/3}$, the solution of the qKZ equation is also
an eigenvector of the transfer matrix of the dense $O(1)$ loop model
with open boundary conditions. Let us recall this fact here
briefly. Let us consider the lattice in the following picture
\sk
\begin{center}
\includegraphics[scale=.6]{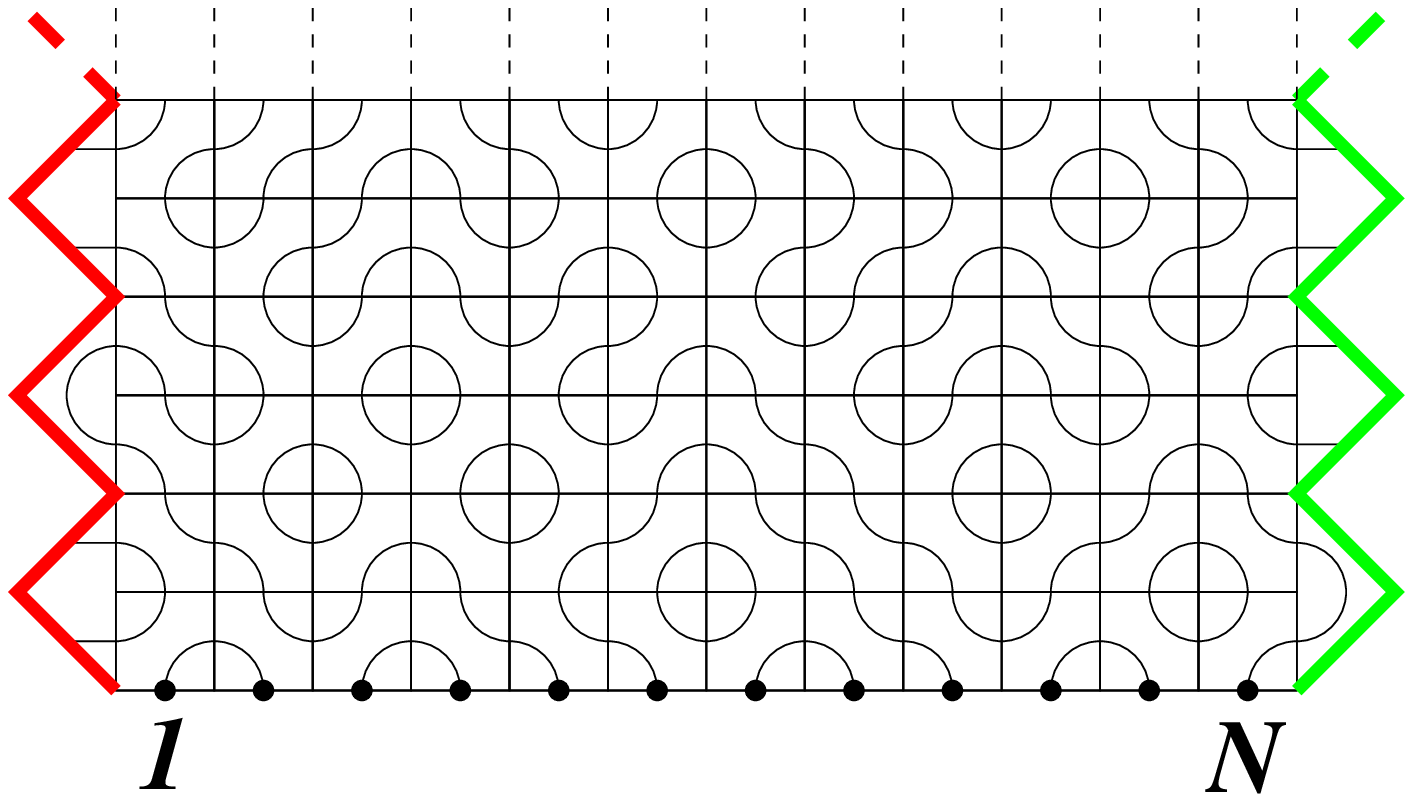}
\end{center}
\noindent
the strip has horizontal length $N$ and vertically it extends to
infinity. 
Each face in the bulk of the lattice is filled with the following
configurations 
\begin{center}
\includegraphics[scale=.6]{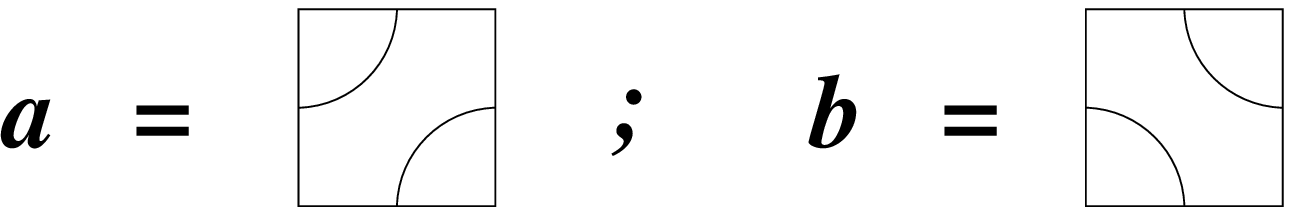}
\end{center}
\noindent
with probabilities $a$ and $b=1-a$, 
while the boundary faces can be in the following configurations 
\begin{center}
\includegraphics[scale=.6]{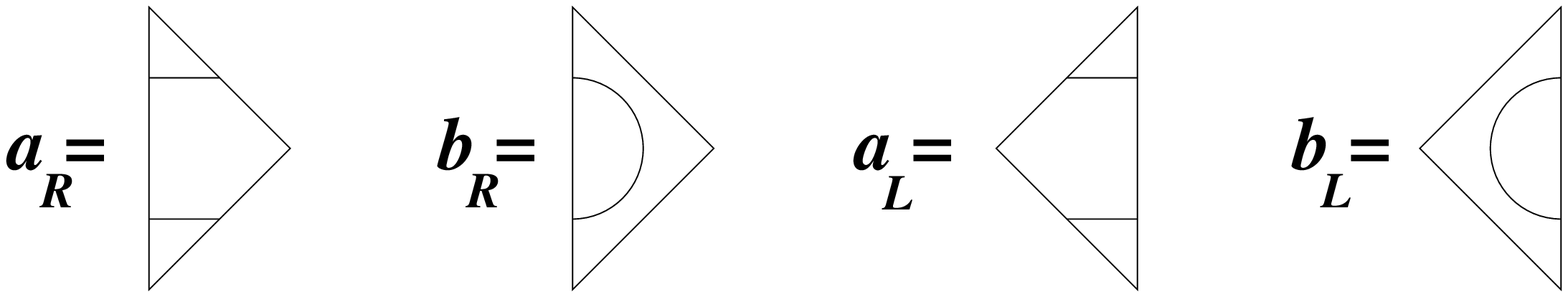}
\end{center}
\noindent
with probability $a_R$,  $b_R=1-a_R$,  $a_L$ and  $b_l=1-a_L$.
The end points of the lines on the bottom boundary are labelled from
$1$ to $N$. 
We are interested in the probability of the connectivity patterns of
such boundary points: a point $i$ is connected by a line either to
another point $j$ or to the left boundary or to the right
boundary. Therefore these connectivity patterns are encoded by the 
extended link patterns defined in the Section \ref{extended }. The
standard way 
to deal with the question of finding the probability of a link pattern
is trough a transfer matrix approach. If one adds to the bottom
a further double row as in the picture 
\begin{center}
\includegraphics[scale=.6]{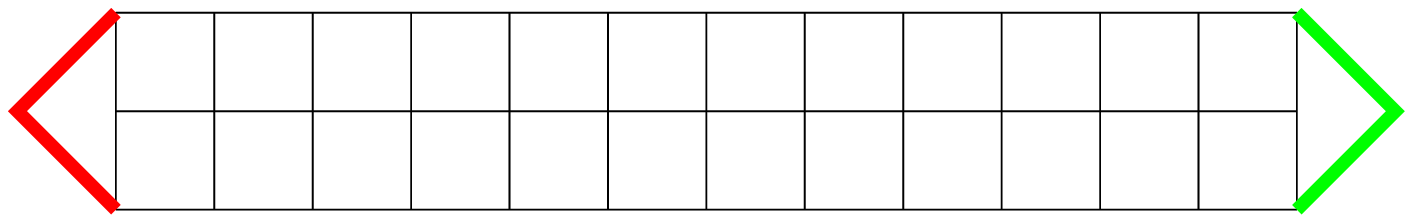}
\end{center}
\noindent
the probabilities of the different link patterns must be
preserved. This means that the vector $\Psi_N\in \sH_N^{(oo)}$ of
probabilities is an eigenvector of the double row transfer matrix
defined above, with eigenvalue equal to $1$. One can show the
integrability of this model by constructing a family of commuting
transfer matrices $T(t)$, depending on a ``horizontal'' spectral
parameter. For this one need to introduce introduce the R and K
matrices, which parametrise the probabilities by 
means of the spectral parameters, as in done for example in
\cite{pdf-pzj-1}. 
The existence of a family of commuting transfer matrices is a
consequence of the Yang-Baxter and Boundary Yang-Baxter equations.
This remains true even if we consider different bulk and boundary
spectral parameters on 
each vertical spectral line \cite{pdf-pzj-1} and the double row
transfer matrix depends on such parameters: $T(t|\zeta_L, \vec
z,\dots, \zeta_R)$. 
When we restrict to the homogeneous problem (all the vertical spectral
parameters equal to $1$) it is not difficult to show that the
following hamiltonian  
$$
H^{(\alpha_L, \alpha_R)}_N = \sum_{i=1}^{N-1} (e_i-1) + \alpha_R(f_R-1) +
\alpha_L(f_L-1),
$$
where $\alpha_L= \frac{1}{\zeta_L+\zeta_L^{-1}+1}$ and $\alpha_R=
\frac{1}{\zeta_R+\zeta_R^{-1}+1}$, commutes 
with the transfer matrix. This is consequence of the fact 
that apart from  a constant term one has 
$$
H^{(\alpha_L, \alpha_R)}_N = T^{-1}(0)\frac{d}{dt}T(t)|_{t=0}.
$$

The vector of probabilities is an eigenvector of $H^{(\alpha_L,
\alpha_R)}_N$ with eigenvalue equal to zero, i.e. is the stationary 
measure of the stochastic matrix $H^{(\alpha_L, \alpha_R)}_N$. 
Closed boundary conditions, for example to the right, are obtained by
sending the boundary spectral parameter $\zeta_R\rightarrow
\infty$. If we consider systems with mixed boundary conditions
(closed-open or open closed) or with closed BCs on both sides we
can perform also only a partial homogeneous specialization. For example
if we have open BCs on the right and closed on the left we can take all
the spectral parameters equal to 1 except the boundary one and the
last bulk one. The corresponding
hamiltonian (i.e. the logarithmic derivative of the transfer matrix in
$t=1$) assumes again a simple form
$$
H^{(c,o)(\alpha_1, \alpha_R)}_N = \sum_{i=2}^{N-1} (e_i-1)   + \alpha_1(e_1-1)+
\alpha_R(f_R-1)
$$
where $\alpha_1=\frac{1}{z_1+z_1)^{-1}+1}$. 
This remark will be important in the following when we will prove
certain mappings between the vectors $\Psi_N$ with different boundary
conditions and different size.

The general problem  of finding the stationary measure in presence of
different spectral parameter is strictly related to the
qKZ equations (see \cite{pdf-pzj-1} for an extended discussion). Indeed the
eigenvector of the transfer matrix $\Psi_N(\zeta_L, \vec z, \zeta_R)$
can be normalized in such a way it 
is a polynomial in the spectral parameters. Moreover as a
simple consequence of the YBE and of the BYBE, at
$q^{1/2}=e^{-2\pi i/3}$ we have
\begin{equation}
T(t|\zeta_L, \dots z_i,z_{i+1},\dots, \zeta_R)\check R_i(z_i,z_{i+1})=
\check R_i(z_i,z_{i+1}) T(t|\zeta_L, \dots z_{i+1},z_{i},\dots, \zeta_R)
\end{equation}
\begin{equation}
T(t|\zeta_L, \dots, 1/z_N, \zeta_R)K_R(\zeta_R, z_N)= K_R(\zeta_R,
z_N)T(t|\zeta_L, \dots, z_N, \zeta_R) 
\end{equation}
\begin{equation}
T(t|\zeta_L, z_1,\dots, \zeta_R)K_L(\zeta_L, z_1)= K_L(\zeta_L,
z_1)T(t|\zeta_L, z_1,\dots, \zeta_R) .
\end{equation}
This means that (by slightly changing the normalization, which makes
it a Laurent polynomial) $\Psi_N(\zeta_L, \vec z, \zeta_R)$ satisfies
the qKZ equations. Notice that at $q^{1/2}=e^{-2\pi i/3}$ all the loops
have weight $1$.

\subsection{Norm of $\Psi_N(\zeta_L, \vec z, \zeta_R)$}

At the special point $q^{1/2}=e^{-2\pi i/3}$ we can compute the sum
rule for the components in the basis of extended link patterns. At
this special point each operator $e_i$, $f_R$ and $f_L$ has an
eigenvector with eigenvalue $1$ in the dual representation  on
$\sH_N^{(oo)*}$, given by 
$\langle v|=\sum_{\pi} \langle \pi|$, where the sum extends to all the the link
patterns. This means that we have
\begin{equation}\label{lefteigen}
\langle v|\check R_i(z_{i+1},z_i)=\langle v|,~~~~~\langle
v|K_L(\zeta_L,z_1)=\langle v|~~~~~ \langle v|K_R(\zeta_R,z_N)=\langle
v|. 
\end{equation}
The sum of the components of $\Psi_N$ is given by
\begin{equation}
\textrm{Sum}_N(\zeta_L;\vec z; \zeta_R) := \sum_\pi
\Psi_{N,\pi}(\zeta_L;\vec z; \zeta_R) = \langle
v|\Psi_{N}(\zeta_L;\vec z; \zeta_R)\rangle 
\end{equation}
As a consequence of the qKZ equation and of eq.(\ref{lefteigen}) we have
$$
\textrm{Sum}_N(\zeta_L;\dots z_i, z_{i+1}, \dots; \zeta_R)= \langle
v|\check R_i(z_{i+1},z_i)|\Psi_{N}(\zeta_L; \dots z_i, z_{i+1}, \dots;
\zeta_R)\rangle 
$$
\begin{equation}
=\langle
v|\Psi_{N}(\zeta_L; \dots z_{i+1}, z_{i}, \dots;
\zeta_R)\rangle = \textrm{Sum}_N(\zeta_L;\dots z_{i+1}, z_{i}, \dots; \zeta_R)
\end{equation}
and analogously
\begin{equation}
\textrm{Sum}_N(\zeta_L;z_1,\dots, z_N;
\zeta_R)=\textrm{Sum}_N(\zeta_L;1/z_1,\dots, z_N; \zeta_R)=
\textrm{Sum}_N(\zeta_L;z_1,\dots, 1/z_N; \zeta_R)
\end{equation}
Therefore $\textrm{Sum}_N(\zeta_L;\vec z; \zeta_R)$ is a symmetric
Laurent polynomial invariant under $z_i\rightarrow 1/z_i$ of degree
width $4N-2$ in each variable,hence it determined by the values at
$4N-1$ distinct values of say $z_N$. 
The bulk recursion relations (combined 
with the symmetry of the polynomial under exchange $z_i\leftrightarrow
z_{i+1}$) provide us with the value of $\textrm{Sum}_N(\zeta_L;z_1,\dots, z_N;
\zeta_R)$ at $z_N= q^{\pm 2\pi i /3} z_j^{\pm 1}$, which means only at $4(N-1)$
points. 
But this is not enough to construct
$\textrm{Sum}_N(\zeta_L;\vec z; \zeta_R)$ by Lagrange
interpolation. 
Moreover it is also possible that at $q^{1/2}=e^{-2\pi i/3}$ there could be some
accidental simplifications leading to the appearance of a common
factor in  the solution
of qKZ for $q$ generic. Such a common
factor should be canceled if one consider the lowest degree
solution.  
The aim of the following sections is to explain how to determine 
$\textrm{Sum}_N(\zeta_L;\vec z; \zeta_R)$ following a different path
and will be the 
consequence of a stronger result. The idea will be to reconstruct the
full vector $\Psi_{N}(\zeta_L;\vec z; \zeta_R)$ from the knowledge of
the solution of the qKZ equations with a closed boundary conditions on
the left and open on the right. For this we need to construct certain
intertwining maps which is done in the next subsection. 

\subsection{Quotients}\label{projection}


From its definition one can think for example at the algebra 
$TL^{(o,c)}_N$ as a subalgebra of $TL^{(o,o)}_N$. However in the case
$\tau=\tau_L=\tau_R=\tau_c=1$ we can think at it as a quotient of
$TL^{(o,o)}_N$ or of $TL^{(o,o)}_{N-1}$. Indeed one obtains
$TL^{(o,c)}_N$ from $TL^{(o,o)}_N$ by quotienting the relation
$f_R=1$, or from $TL^{(o,o)}_{N-1}$ by relabelling $e_{N-1}=f_R$ and
quotienting over the missing relation
i.e. $e_{N-1}e_{N-2}e_{N-1}=e_{N-1}$.  
In the present section we want to implement this idea at the level of
representations. We construct a map:
$\Upsilon_N : \sH_N^{(oo)}\rightarrow \sH^{(oc)}_{N}$,
which intertwines between the representation of $TL^{(o,c)}_N$ and the
one of $TL^{(o,o)}_N$ with
\begin{equation}\label{quoz1}
\Upsilon_N e_{i<N}= e_{i<N} \Upsilon_N,~~~ \Upsilon_N f_L= f_L \Upsilon_N,~~~
\Upsilon_N f_R =  \Upsilon_N.   
\end{equation}

At a graphical level the map $\Upsilon_N$ is obtained as follows:
starting from the right remove the first two lines connected to $R$
and connect the unmatched points with an arc.
If the initial number of lines emanating from $R$ was even, repeat
this procedure with the remeaning lines joining $R$ until the lines
from $R$ are exhausted. If the number of lines from $R$ was odd repeat
the procedure until there is a single line connected to $R$ then
remove this line and connect the unmatched bulk point with the left
boundary $L$. Here is an example
\sk
\begin{center}
\includegraphics[scale=.6]{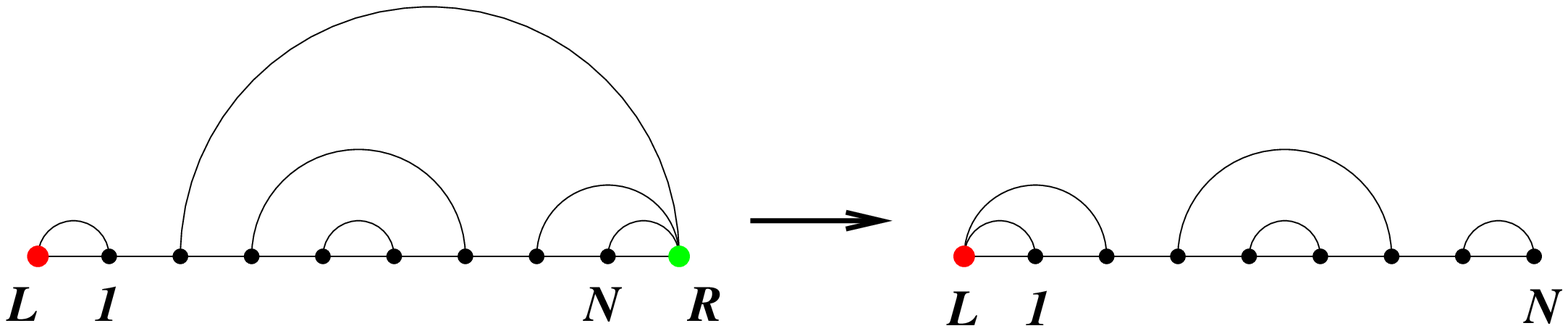}
\end{center}
\sk
The map
$\Upsilon_N$ is easier to visualize if we use a different diagrammatic
representation of the components, in terms of doubly-blobbed arcs
\cite{JS2}, then $\Upsilon_N$ simply consists in removing the right
blobs.  
The other map we want to define is $\Theta_N : \sH_N^{(oo)}\rightarrow
\sH^{(oc)}_{N+1}$, which intertwines $f_R$ and $e_{N-1}$
\begin{equation}\label{quoz2}
\Theta_N e_{i<N}= e_{i<N} \Theta_N,~~~ \Theta_N f_L= f_L
\Theta_N,~~~\Theta_N f_R = e_{N} \Theta_N. 
\end{equation}
At the diagrammatic level the map $\Theta_N$ goes as follows: change
the name of the the point the right boundary point $R\rightarrow N+1$.
If there are no lines connected to $N+1$ draw a line from it to $L$.
If there are more than two lines connected to $N+1$ remove the
second and the third one starting from the right and draw an arc
connecting the unmatched bulk points.
Repeat this procedure until there are either one or two remaining
lines. In the former case do nothing, while in the latter case remove
the left most line and connect the unmatched bulk point with the left
boundary $L$. Here is an example
\sk
\begin{center}
\includegraphics[scale=.6]{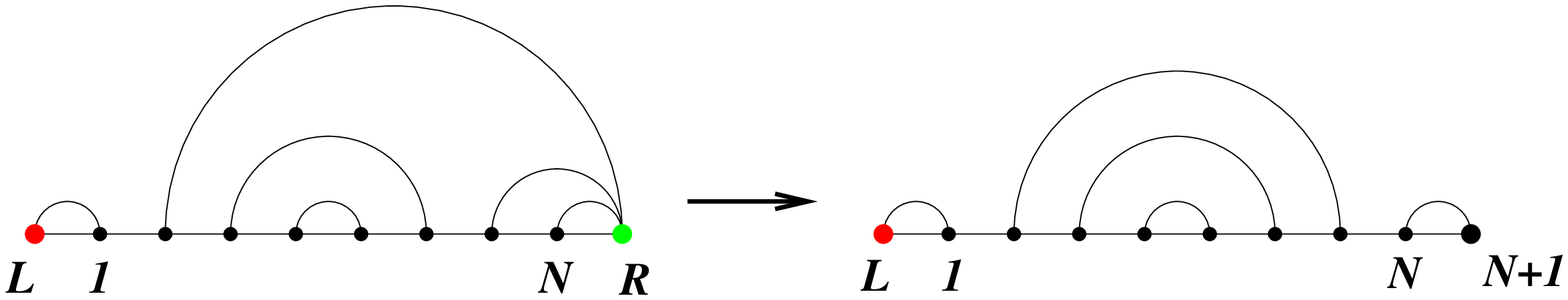}
\end{center}
\sk
The proof of eqs.(\ref{quoz1}, \ref{quoz2}) is a tedious but
strightforward graphical case by
case analysis.

\subsection{Reconstruction of  $\Psi_{N}(\zeta_L;\vec z; \zeta_R)$}\label{reconstruction}

Now we use the maps $\Upsilon_N$ and $\Theta_N$ to reconstruct
$\Psi_{N}(\zeta_L;\vec z; \zeta_R)$ from the known form of the
solution of qKZ with mixed boundary conditions \cite{paul} and gain
information on the sum rule.
First of all we notice that (with a slight abuse of notation)
\begin{equation}
\Upsilon_N K_L(\zeta_L| z) = K_L(\zeta_L| z)\Upsilon_N,~~~~
\Upsilon_N \check R_i(z,w)=  R_i(z,w)\Upsilon_N,~~~~~
\Upsilon_N K_R(\zeta_R| z) =\Upsilon_N,
\end{equation}
therefore we find that $\Upsilon |\Psi_N(\zeta_L,\vec z, \zeta_R)\rangle$
satisfies the qKZ equations with mixed closed-open boundary
conditions. Therefore it must be 
\begin{equation}
\Upsilon |\Psi_N(\zeta_L,\vec z, \zeta_R)\rangle = f_1(\zeta_L,\vec z,
\zeta_R) |\Psi_N^{(o,c)}(\zeta_L,\vec z)\rangle
\end{equation}
where $f_1(\zeta_L,\vec z,\zeta_R)$ is a symmetric polynomials in the
$z_i$s invariant under $z_I\leftrightarrow 1/z_i$ and
$|\Psi_N^{(o,c)}(\zeta_L,\vec z)\rangle$  is the known
solution of the qKZ equations with mixed B.C. \cite{paul}. Moreover
since we have $\langle v| = \langle v_N^{(o,c)}|\Upsilon$, where
$\langle v_N^{(o,c)}| = \sum_{\pi^{(o,c)}} \langle \pi^{(o,c)}| $ and
the sum runs over all right extended link patterns, we obtain for the
sum rule
$$
\langle v|\Psi_N(\zeta_L,\vec z, \zeta_R)\rangle = f_1(\zeta_L,\vec
z,\zeta_R) \langle v^{(o,c)}|\Psi_N^{(o,c)}(\zeta_L,\vec z, \zeta_R)\rangle.
$$
In \cite{paul} it is proven that $\langle
v_N^{(o,c)}|\Psi_N^{(o,c)}(\zeta_L,\vec z)\rangle = \chi_{N-1}(\vec
z)\chi_{N}(\vec z, \zeta_L)$, this means that we have found two 
factors of the sum rule
\begin{equation}\label{part_sum1}
\textrm{Sum}_N(\zeta_L;\vec z; \zeta_R)\propto \chi_{N-1}(\vec
z)\chi_{N}(\vec z, \zeta_L), 
\end{equation}
where $\chi_{n}(z_1,\dots, z_n)$ is the symplectic character
\begin{equation}
\chi_{n}(z_1,\dots, z_n) =\frac{\det\left(z_i^{j+ \lceil j/2 \rceil -1}- z_i^{-j -
      \lceil j/2 \rceil +1}\right)_{1\leq i,j, \leq n}}{\det\left(z_i^{j}-
    z_i^{-j}\right)_{1\leq i,j, \leq n}}. 
\end{equation}
In order to find the other factors, we use the map $\Theta_N$. As done 
above with $\Upsilon$ one can prove that
\begin{equation}
\Theta_N |\Psi_N(\zeta_L,\vec z, \zeta_R)\rangle = f_2(\zeta_L,\vec
z,\zeta_R) |\Psi_{N+1}^{(o,c)}(\zeta_L;\vec z, \zeta_R)\rangle
\end{equation}
where now $|\Psi_{N+1}^{(o,c)}(\zeta_L;\vec z, \zeta_R)\rangle$ is
solution of the qKZ equations with mixed boundary conditions at size
$N+1$ and the parameter 
$\zeta_R$ plays the role of a bulk spectral parameter. We have also
$\langle v| = \langle v_{N+1}^{(o,c)}|\Theta_N$, and using again the
results of \cite{paul} we obtain for the sum rule
\begin{equation}\label{part_sum2}
\textrm{Sum}_N(\zeta_L;\vec z; \zeta_R)\propto \chi_{N}(\vec z,
\zeta_R)\chi_{N+1}(\vec z,  \zeta_L, 
\zeta_R).
\end{equation}
To conclude the derivation we notice that these four factors exhaust
the degree of $\textrm{Sum}_N(\zeta_L;\vec z; \zeta_R)$ (and do not
have common factors), therefore 
\begin{equation}\label{resultsum1}
\textrm{Sum}_N(\zeta_L;\vec z; \zeta_R)=\chi_{N}(\vec
z)\chi_{N+1}(\zeta_L, \vec z)\chi_{N+1}(\zeta_R, \vec
z)\chi_{N+2}(\zeta_R, \zeta_L, \vec z).
\end{equation}
Let us notice that eqs.(\ref{part_sum1},\ref{part_sum2}) provides us
also with a way of finding all the components of $|\Psi_N(\zeta_L,\vec
z, \zeta_R)\rangle$ without the need of solving the qKZ equations for
generic $q$. Let us show this in a
concrete example at $N=3$. Component wise
eq.(\ref{part_sum1},\ref{part_sum2}) read  
$$
\psi_{ccc} +\psi_{cco} = f_1 \psi_{ccc^{(o,c)}},~~\psi_{occ}+\psi_{oco}= f_1
\psi_{occ}^{(o,c)} 
$$
$$
\psi_{coc}+\psi_{coo}+\psi_{ooo}+\psi_{ooc}= f_1 \psi_{coc}^{(o,c)}
$$
$$
\psi_{ccc} = f_2 \psi_{cccc}^{(o,c)},~~\psi_{coc} = f_2
\psi_{cocc}^{(o,c)},~~\psi_{occ} = f_2 \psi_{occc}^{(o,c)},~~ \psi_{ooc} = f_2 \psi_{oocc}^{(o,c)},~~
$$
$$
\psi_{coo}+\psi_{cco}= f_2 \psi_{ccoc}^{(o,c)},~~\psi_{ooo}+\psi_{oco}=f_2
\psi_{ococ}^{(o,c)}. 
$$
This system of equations is immediately solved
$$
\psi_{ccc} = f_2 \psi_{cccc}^{(o,c)},~~\psi_{coc} = f_2
\psi_{cocc}^{(o,c)},~~\psi_{occ} = f_2 \psi_{occc}^{(o,c)},~~ \psi_{ooc} = f_2 \psi_{oocc}^{(o,c)}
$$
$$
\psi_{cco} = f_1 \psi_{ccc}^{(o,c)} -f_2\psi_{cccc}^{(o,c)},~~ \psi_{oco} = f_1
\psi_{occ}^{(o,c)} -f_2\psi_{occc}^{(o,c)},~~ 
$$
$$
\psi_{coo}= f_2(\psi_{cccc}^{(o,c)}+\psi_{ccoc}^{(o,c)})-f_1\psi_{ccc}^{(o,c)},~~\psi_{ccc}=
f_2(\psi_{occc}^{(o,c)}+\psi_{ococ}^{(o,c)})-f_1\psi_{occ}^{(o,c)}. 
$$
giving an alternative way of computing the components of the ground
state of the dense $O(1)$ model with open boundary conditions. We do
not have a proof that this method allows to reconstruct the full
ground state for a generic system size $N$, but we have checked it up
to $N=6$, checking as well that the result coincide with the ones
found by solving the qKZ equation and then taking the limit $q^{1/2}\rightarrow
e^{-2\pi i/3}$. 

\section{Conclusions and perspectives}

In this paper we have considered the qKZ equations related to the
baxterization of the two boundaries Temperley Lieb algebra. By
deriving a recursion relation satisfied by their solution we have been
able to  construct it explicitly. At the combinatorial
point $q^{1/2}=e^{-2pi i/3}$ the solution of the qKZ equations is also
the ground state of the dense $O(1)$ loop model with open boundary
conditions. We have shown how to reconstruct this ground state
starting from the ground state of the same model with mixed boundary
conditions known from \cite{paul}. In doing that we have also been
able to prove the sum rule. In Section \ref{affine_sectio} we have briefly
discussed the relations between our problem and the theory of
representation of (doubly) affine Hecke algebras of type
$C\vee C_N$. It would be interesting to develop further this
observation, in particular making contact with a recent work of
Kasatani \cite{kasatani} and hopefully with the theory of non symmetric
Koornwinder-MacDonald polynomials. It would be also interesting to
find integral formulae as already done in the case of other boundary
conditions \cite{conjecture}
In this paper we have only tangentially touched problems related to
the combinatorics behind the homogeneous limit of the ground state of
the $O(1)$ model. In facts it is not difficult to see that, making use of
the mapping between model with different boundary conditions, one can
easily explain certain observation reported in Table 2 of
\cite{mitra}. We plan to come back to such matters and to
generalization to even more exotic boundary conditions in the immediate
future. 

\section*{Acknowledgments}

It is a pleasure for us to thank P. Zinn-Justin for many fruitful
discussions. We thank also the LPTMS (UMR 8626 du CNRS)
Universit\'e Paris Sud, where part of this work was done. We
acknowledge the financial support of the ANR 
programs ``GIMP'' ANR-05-BLAN-0029-01 and ``SLE'' ANR-06-BLAN-0058-01.

\appendix

\section{Recursion relations: boundary}\label{recursion_bound}

In this appendix we want to derive a boundary recursion relation
similar to the bulk one, the motivation being twofold. On one side we
show that leaving the boundary weight $\tau_R$ free, the boundary recursion
relation maps the solution of qKZ at size $N$ to a solution of qKZ at
size $N-1$ and $\tau_R\rightarrow 1/\tau_R$. This provides a
motivation to restrict to the case $\tau_R=1$ (and similarly
$\tau_L=1$). On the other side the recursion we find provides the most
convenient way to write the Laplace interpolation formula giving the
unknown factor $\phi^{(R)}_N$ of $\psi\underbrace{_{oo\dots oo}}_N$.
From the analysis of Section \ref{trivial_factors} adapted to the case
$\tau_R\neq 1$, it follows that if $z_N= 
q/\zeta_R$ or $z_N = \zeta_R/(qk) $ then all the components in which the
point $N$ is not 
connected to the right boundary $R$ are zero. The other components,
namely the ones having an arc going from $N$ to $R$ are in one to one
correspondence with the extended link patterns with one bulk point
removed. One simply removes the point $N$ and the arc from it to
$R$. We call this map $p_R:\sH_{N}^{(oo)}\rightarrow
\sH_{N-1}^{(oo)}$. In this section we show that 
$p_R\circ\Psi_N(\dots,z_{N-1},z_{N}=q/\zeta_R ~\textrm{ or}~z_N=
\zeta_R/(qk);\zeta_R)$ satisfy a 
modified qKZ equation. Let us restrict ourselves to the case $z_N=
q/\zeta_R$, the other one being completely analogous. First of all we
notice that 
$p_R\circ\Psi_N(\dots,z_{N-1},z_{N}=q/\zeta_R;\zeta_R)$ satisfies the qKZ
eqs.(\ref{qKZ1}) for $i< N-1$ and eq.(\ref{qKZ3}), these properties
being a trivial consequence of the fact that
$\Psi_N(\dots,z_{N-1},z_{N}=q/\zeta_R;\zeta_R)$ satisfies the same
equation.
Then let us consider the following operator 
$$
\tilde K_R(\zeta_R|z) = R_{N-1}(1/x,q/\zeta_R)K_R(\zeta_R|z)R_{N-1}(q/\zeta_R,x).
$$ 
We see easily that
\begin{equation}
\tilde K_R(\zeta_R|z_{N-1})\Psi_N(\dots,z_{N-1},z_{N}=q/\zeta_R;\zeta_R) =
\Psi_N(\dots,1/z_{N-1},z_{N}=q/\zeta_R;\zeta_R) 
\end{equation}
As one expected, $\tilde K_R(z_{N-1})$
preserves the image of $f_R$ (which are the components not identically zero when
$z_N=q/\zeta_R$) therefore if we let it act only on such a subspace we
see that it assumes the following form 
\begin{equation}
\tilde K_R(\zeta_R,z_{N-1}) = \frac{(z\tilde \zeta_R-q^4)(z \tilde k-
  \tilde \zeta_R)}{(\tilde \zeta_R-q^4 z)(\tilde k-
  \tilde \zeta_R z)}\left(\frac{(z-\tilde \zeta_R)(z-\tilde k/ 
    \tilde \zeta_R)I +
  \frac{(q^2-1)}{(q + \tilde \tau_R)}(z^2-1)\tilde
  f_R}{(z\tilde\zeta_R-1)(z\tilde k/\tilde \zeta_R-1)}\right)
\end{equation}
where $\tilde f_R = 1/\tau_R f_R e_{N-1}$, $\tilde \zeta_R=
q^3/\zeta_R$, $\tilde \tau_R=1/\tau_R$ and $\tilde k$ is the same as $k$
in which we substitute $\tau_R$ with $\tilde \tau_R$. Apart from the
multiplicative 
factor in front of it, this has just the 
same form of the scattering matrix in eq.(\ref{boundary_scatt}). This
is consistent 
with the fact that the algebra of operators
$f_L$, $e_{i<N-1}$ and $\tilde f_R$ is a $TL_{N-1}^{(o,o)}$ with
boundary loop weight $\tilde
\tau_R$. From now on we will restrict ourselves to
the case when $\tau_R=\tau_L=1$. This implies in particular $\tilde k=
k=q$ and we can write 
\begin{equation}
\tilde K_R(\zeta_R| z)=  \frac{(z-q \zeta_R)(z -
  q^2/ \zeta_R)}{(1-q z \zeta_R)(1 -
  z q^2/ \zeta_R)} K_R(q^3/\zeta_R| z)
\end{equation}
We conclude therefore that
$p_R\circ\Psi_N(\dots,z_{N-1},z_{N}=q/\zeta_R;\zeta_R)$ satisfies  the
following set of equations
\eq
\check{R}_{i<N-1}(z_{i+1},z_i) p_R\circ\Psi_N(\dots, z_i,
z_{i+1},\dots, z_N=q/\zeta_R) = 
p_R\circ\Psi(z_{i+1}, z_{i},\dots, z_N=q/\zeta_R);
\en
\eq
\frac{u_{R}(1/z_{N-1})}{u_{R}(z_{N-1})} K_R(q^3/\zeta_R|z_{N-1})
p_R\circ\Psi_N(\dots,z_{N-1},z_{N}=q/\zeta_R;\zeta_R) =
p_R\circ\Psi_N(\dots,1/z_{N-1},z_{N}=q/\zeta_R;\zeta_R); 
\en
\eq
K_L(\zeta_L|z_1)
p_R\circ\Psi_N(\zeta_L;z_1,\dots,z_{N}=q/\zeta_R;\zeta_R) =
p_R\circ\Psi_N(\zeta_L,s/z_1\dots,z_{N}=q/\zeta_R;\zeta_R); 
\en
where $u_R(z)=\left(\frac{1}{q z}-\zeta_R\right)\left(q z
  -\frac{\zeta_R}{q}\right)$. Now let us suppose that 
we can find a function $g_R(z)$ which is invariant under $z\rightarrow
1/z$ and such that
\begin{equation}\label{rel_rec1}
\frac{u_R(z)g_R(z)}{u_R\left(\frac{s}{z}\right)g_R\left(\frac{s}{z}\right)}=
\frac{\left(\frac{1}{q    
      z}-\zeta_R\right)\left(q 
  z-\frac{\zeta_R}{q}\right)}{\left(\frac{z}{s q}-\zeta_R\right)\left(\frac{s q }{z}
    -\frac{\zeta_R}{q} \right)}\frac{g_R(z)}{g_R\left(\frac{s}{z}\right)}=1  
\end{equation}
then it is straightforward to see that 
\begin{equation}\label{rec1}
\prod_{i=1}^{N-1} \left(\frac{1}{u_R(z_i)g_R(z_i)} \right)
p_R\circ\Psi_N(\dots,z_{N-1},z_{N}=q/\zeta_R;\zeta_R)
\end{equation}
is solution of the unmodified qKZ equation with boundary parameter
$\tilde \zeta_R=q^3/\zeta_R$.
If we require that the solution $g(z)$ of eq.(\ref{rel_rec1}) is a
Laurent polynomial in $z$, then we find that the parameter $s$ must be of the
form $s=q^{-3/n}$ and  it is easy to check that the solution is
\begin{equation}
g(z)=\prod_{j=1}^{n-1}(q^{3j/n-1}z-\zeta_R)(q^{3j/n-1}/z- \zeta_R),
\end{equation}
therefore, calling 
$$
G_R(z_1,\dots, z_{N-1}) = \prod_{i=1}^{N-1} u_R(z_i)g_R(z_i) \propto
\prod_{i=1}^{N-1}\prod_{j=1}^{n}(q^{3j/n-1}z_i-\zeta_R)(q^{3(j-1)/n-1}/z_i-
\zeta_R)  
$$
we can rewrite eq.(\ref{rec1}) as follows
\begin{equation}\label{recursion}
p_R\circ\Psi_N(\dots,z_{N-1},z_{N}=q/\zeta_R;\zeta_R) =
k_R(\zeta_L, \zeta_R)G_R(z_1,\dots, z_{N-1})\Psi_{N-1}(\dots,z_{N-1};q^2/\zeta_R)
\end{equation}
where $k_R(\zeta_L, \zeta_R)$ is an unknown function of $\zeta_L$
and $\zeta_R$, that we can fix  for $s=q^{-3/2}$ to be$k_R(\zeta_L,
\zeta_R)= \left(s\zeta_R-\frac{q^2}{\zeta_L}\right) 
\left(\frac{\zeta_L}{\zeta_R}-q\right),$ by using minimality
arguments similar to the ones 
employed in Section \ref{sol_section}. Now we want to change a bit
perspective on the role of the boundary parameter $\zeta_R$; as a
consequence of the bulk recursion relations, it follows that the
function $\phi^{R}_N(\zeta_L;\vec z;\zeta_R)$ is a Laurent polynomial
in $\zeta_R$, invariant under $\zeta_R\rightarrow q^3/\zeta_R$ and
degree width $\lfloor\frac{N-1}{2} \rfloor$. This
means that we can use the boundary recursion relation in order to
construct $\phi^{R}_N$ by using Lagrange interpolation in
$\zeta_R$ and the symmetries of $\phi^{R}_N$ under exchange of the
$z_i$s. Forgetting a constant normalization factor that we cannot fix,
the result reads
\begin{equation}
\phi^{(R)}_N(\zeta_L;\vec z;\zeta_R) = \sum_{i=1}^N\prod_{k\neq i}
\frac{\left(\zeta_R+\frac{q^3}{\zeta_R}-q^2z_k-\frac{q}{z_k} 
  \right)\left(z_i +\frac{q}{z_i}-q^2 z_k -\frac{1}{qz_k}
  \right)}{\left(q^2z_i+\frac{q}{z_i}  
    -q^2z_k-\frac{q}{z_k}\right)} \phi^{(R)}_{N-1}(\zeta_L;\hat z_i;q/z_i)
\end{equation}
It is remarkable that this expression turn out not only to be a 
Laurent polynomial in the spectral parameters $z_i$, but it is also
endowed with all the symmetry properties expected for
$\phi^{(R)}_{N}$. We have checked for that for systems up to size
$N=6$, starting from the above expression one gets a solution of the
qKZ equations.

%
%
%


\end{document}